\documentclass[11pt,preprint]{aastex}

\slugcomment{submitted to ApJ}
\shorttitle{Momentum Diffusion of Radiating Electrons}
\shortauthors{Stawarz \& Petrosian}

\begin{document}

\title{On the Momentum Diffusion of Radiating Ultrarelativistic Electrons in a
Turbulent Magnetic Field}

\author{\L ukasz Stawarz\altaffilmark{1, 2} and Vahe Petrosian\altaffilmark{3}}
\email{stawarz@slac.stanford.edu}
\altaffiltext{1}{Kavli Institute for Particle Astrophysics and Cosmology, Stanford
University, Stanford CA 94305}
\altaffiltext{2}{Astronomical Observatory, Jagiellonian University, ul. Orla 171, 30-244
Krak\'ow, Poland}
\altaffiltext{3}{Center for Space Science and Astrophysics, Department of Physics and Applied
Physics, Stanford University, Stanford, CA 94305}

\begin{abstract}
Here we investigate some aspects of stochastic acceleration of 
ultrarelativistic electrons by magnetic turbulence. In particular, we 
discuss the steady-state energy spectra of particles undergoing momentum 
diffusion due to resonant interactions with turbulent MHD modes, taking 
rigorously into account direct energy losses connected with different 
radiative cooling processes. 
For the magnetic turbulence we assume a given power spectrum 
of the type $\mathcal{W}(k) \propto k^{-q}$.
In contrast to the previous approaches, however, we assume a finite range of 
turbulent wavevectors $k$, consider a variety of turbulence spectral 
indexes $1 \leq q \leq 2$, and concentrate on the case of a very 
inefficient particle escape from the acceleration site. We find that for 
different cooling and injection conditions, stochastic acceleration 
processes tend to establish a modified ultrarelativistic Maxwellian 
distribution of radiating particles, with the high-energy exponential 
cut-off shaped by the interplay between cooling and acceleration rates. 
For example, if the timescale for the dominant radiative process scales 
with the electron momentum as $\propto p^r$, the resulting electron energy 
distribution is of the form $n_{\rm e}(p) \propto p^2 \, \exp\left[ - {1 
\over a} \, \left(p / p_{\rm eq}\right)^a\right]$, where $a = 2-q-r$, and 
$p_{\rm eq}$ is the equilibrium momentum defined by the balance between 
stochastic acceleration and energy losses timescales. We also discuss in 
more detail the synchrotron and inverse-Compton emission spectra produced by 
such an electron energy distribution, taking into account Klein-Nishina 
effects. We point out that the curvature of the high frequency segments of 
these spectra, even though being produced by the same population
of electrons, may be substantially different between the synchrotron 
and inverse-Compton components.
\end{abstract}


\section{Introduction}

Stochastic acceleration of ultrarelativistic particles via scatterings by 
magnetic inhomogeneities was the first process discussed in the context of
generation of a power-law energy distribution of cosmic rays
\citep{fer49,dav56}. Because the characteristic acceleration timescale for a
given velocity of magnetic inhomogeneities, say Alfv{\'e}n velocity $v_A$, is
$t_{\rm acc} \propto (v_{\rm A} / c)^{-2}$, the stochastic particle acceleration
is often referred as a `2nd-order Fermi process'. For commonly occuring
non-relativistic turbulence, $v_{\rm A} \ll c$, turbulent
acceleration mechanism is often deemed less efficient when
compared to  acceleration by shocks where the rate of momentum change $\delta
p/p \sim v_{\rm sh} / c$ (hence the name 1st-order Fermi process). 
However, here one also needs repeated crossing of the shock front by the
particles which can come about via scattering by turbulence upstream and
downstream of the shock. Thus, again the acceleration rate or timescale is
determined by the scattering time scale. For nonrelativistic turbulence $v_A\ll
c$, relativistic particles $p \gg m c^2$, and high-$\beta$ or weakly magnetized plasma, this
time is shorter than the stochastic acceleration time, which may not be the case
in many astrophisical plasmas. We note that in a
relativistic regime, for example, 1st-order Fermi process encounters several
difficulties in accelerating particles to high energies
\citep[e.g.,][]{nie06a,nie06b,lem06}, while at the same time stochastic particle
energization may play a major role, since velocities of the turbulent modes may
be high, $v_{\rm A} \lesssim c$. And indeed, 2nd-order Fermi processes were
being discussed in the context of different astrophysical sources of high energy
radiation and particles, such as accretion discs \citep[e.g.,][]{liu04,liu06},
clusters of galaxies \citep[e.g.,][]{pet01,bru07}, gamma-ray bursts
\citep[e.g.,][]{ste04}, solar flares \citep[e.g.,][]{pet99,pet04}, blazars
\citep[e.g.][]{kat06b,gie07}, or extragalactic large-scale jets
\citep[e.g.,][]{sta02,sta04}. We note, that although turbulent acceleration is often a
process of choice in modeling high energy emission in different objects, and in
fact there may be some other yet much less understood mechanisms responsible for
generation of such (like magnetic reconnection), evidences for the distributed
(or {\it in situ}) acceleration process taking place in several astrophysical
systems are strong \citep[see, e.g.,][in the context of extragalactic
jets]{jes01,ks06,har07}.

It was pointed out by \citet{sch84,sch85}, that continuous (stochastic)
acceleration of high energy electrons undergoing radiative energy losses tends
to establish their ultrarelativistic Maxwellian energy distribution, as long as
particle escape from the acceleration site is inefficient. This analysis
concerned a particular case of acceleration timescale independent on the
electrons' energy, and the dominant synchrotron-type energy losses.
Interestingly, very flat (inverted) electron spectra of the ultrarelativistic
Maxwellian-type --- often approximated as a monoenergetic electron distribution
--- were discussed in the context of flat-spectrum radio emission observed from
Sgr A$^{\star}$ and several active galactic nuclei \citep[see, e.g.,][and
references therein]{bec97,bir01}. More recently, it was proposed that such
`non-standard' electron spectra can account for striking high-energy X-ray
emission of large-scale jets observed by {\it Chandra} satellite
\citep{sta02,sta04}, or correlated X-ray and $\gamma$-ray (TeV) emission from
several BL Lac objects detected by the modern ground-base Cherenkov Telescopes
\citep{kat06a,gie07}. In addition, it was shown that narrow
electron spectra, e.g. Maxwellian
distribution, can explain properties of extragalactic high brightness
temperature radio sources \citep{tsa07a,tsa07b}, alleviating the difficulties
associated with the anticipated by not observed inverse-Compton catastrophe
\citep{ost06}.

Motivated by these most recent observational and theoretical results, in this
paper we investigate further some aspects of stochastic acceleration of
ultrarelativistic electrons by magnetic turbulence. In particular, we discuss
steady-state energy spectra of particles undergoing momentum diffusion due to
resonant interactions with turbulent MHD modes, taking rigorously into account
direct {\it in situ} energy losses connected with different radiative cooling
processes. As described in the next section \S\,2, we use the quasilinear
approximation for the wave-particle interactions, assuming a given power
spectrum $\mathcal{W}(k) \propto k^{-q}$ for magnetic turbulence within some
finite range of turbulent wavevector $k_1<k<k_2$, and consider turbulence
spectral indexes in the range $1 \leq q \leq 2$. In section \S\,3 we provide
steady-state solutions to the momentum diffusion equation corresponding to the
case of no particle escape but different cooling and injection conditions. In
section \S\,4 some particular solutions are given corresponding to the case of a
finite particle escape from the acceleration site. In section \S\,5 we discuss
in more details synchrotron and inverse-Compton emission spectra of
stochastically accelerated electrons, taking into account Klein-Nishina effects.
Final discussion and conclusions are presented in the last section \S\,6 of the
paper.

\section{General Description}

Let us denote the phase space density of ultrarelativistc particles by
$f(\vec{x}, \vec{p}, t)$, such that the total number of particles is
$\mathcal{N}(t) = \int d^3 x \int d^3 p \, f(\vec{x}, \vec{p}, t)$. Here the
position coordinate $\vec{x}$ and the momentum coordinate $\vec{p}$ are not the
position and the momentum of some particular particle, but are fixed to the
chosen coordinate space, and therefore are independent variables. In the case of
collisionless plasma, the function $f(\vec{x}, \vec{p}, t)$ satisfies the
relativistic Vlasov equation with the acceleration term being determined by the
Lorentz force due to the \emph{average} plasma electromagnetic field acting on
particles. This averaged field can be found, in principle, through the Maxwell
equations, and such an approach would lead to the exact description of the
considered system. However, due to strongly non-linear character of the
resulting equations (and therefore substantial complexity of the problem), in
most cases an approximate description is of interest. In the `test particle
approach', for example, one assumes configuration of electromagnetic field and
solves the particle kinetic equation to determine particle spectrum. Further
simplification can be achieved if one assumes presence of only a small-amplitude
turbulence $(\delta \vec{E}, \delta \vec{B})$ in addition to the large-scale
magnetic field\footnote{Due to the expected high conductivity of the plasma one
can neglect the large-scale scale electric field, $\vec{E}_0 = 0$ .}
$\vec{B}_0\gg \delta \vec{B}$, such that the total plasma fields are $\vec{B} =
\vec{B}_0 + \delta \vec{B}$ and $\vec{E} = \delta \vec{E}$.

In order to find the evolution of the particle distribution function in the
phase space under the influence of such fluctuating electromagnetic field, it is
convenient to consider an ensemble of the distribution functions (all equal at
some initial time), such that the appropriate ensemble-averaging gives $\langle
\delta \vec{B} \rangle = \langle \delta \vec{E} \rangle = 0$ and $f(\vec{x},
\vec{p}, t) = \langle f(\vec{x}, \vec{p}, t) \rangle + \delta f(\vec{x},
\vec{p}, t)$. It can be then shown via the `quasilinear approximation' of the
Vlasov equation that the ensemble-average of the distribution function $\langle
f(\vec{x}, \vec{p}, t) \rangle$ satisfies the Fokker-Planck equation
\citep{hal67,mel68}\footnote{The Fokker-Planck equation can be also derived
straight from the definition of the function $f(\vec{x}, \vec{p}, t)$, assuming
that the interaction of the particles with turbulent waves is a Markov process
in which every interaction (collision) changes the particle energy only by a
small amount, and that the recoil of the turbulent modes during the collision
can be neglected \citep{bla87}.}. If, in addition, the particle distribution
function is only slowly varying in space (`diffusion approximation'),
and the scattering time (or mean free path) is shorter than all other relevant
times (or mean free paths), the ensemble-averaged particle distribution function
can be assumed to be spatially uniform and isotropic in $p$, namely $\langle
f(\vec{x}, \vec{p}, t) \rangle = \langle f(p, t) \rangle$, and the Fokker-Planck
equation can be further reduced to the momentum diffusion equation
\citep[see][]{tsy77,mel80,sch02}.

The resulting momentum diffusion equation describing evolution of the particle
distribution can be written as
\begin{equation}
{\partial \over \partial t} \langle f(p, t) \rangle = {1 \over p^2} \, {\partial
\over \partial p} \, \left[ p^2 \, D(p) \, {\partial \over \partial p} \,
\langle f(p, t) \rangle \right] \, ,
\label{diff1}
\end{equation}
\noindent
where the momentum diffusion coefficients $D(p)$ approximates the rate of
interaction with fluctuating electromagnetic field. Several other terms
representing physical process that may influence evolution of the particle
energy spectrum can be added to the diffusion equation (\ref{diff1}). In
particular, one can include continuous energy gains and losses due to direct
acceleration (e.g., by shocks) and radiative cooling. Furthermore, if the
diffusion of particles out of the turbulent region is approximated by a 
catastrophic escape rate (or time $t_{\rm esc}$), and if there is a source term
$\widetilde{Q}(p, t)$ representing particle injection into the system, then the
spatially integrated (over the turbulent region) one-dimensional particle
momentum distribution, $n(p, t) \equiv 4 \pi \, p^2 \, \langle f(p, t) \rangle$,
is obtained from \cite[see, e.g.,][]{pet04}
\begin{equation}
{\partial n(p, t) \over \partial t} = {\partial \over \partial p} \, \left[D(p)
\, {\partial n(p, t) \over \partial p} \right] - {\partial \over \partial p} \,
\left[ \left( {2 \, D(p) \over p} + \langle \dot{p} \rangle \right) \, n(p, t)
\right] - {n(p, t) \over t_{\rm esc}} + \widetilde{Q}(p, t).
\label{diff2}
\end{equation}
\noindent

Let us further assume presence of an isotropic Alfv\'enic turbulence described
by the one-dimensional power spectrum $\mathcal{W}(k) \propto k^{-q}$ with $1
\leq q \leq 2$ in a \emph{finite} wavevector range $k_1 \leq k \leq k_2$, such
that the turbulence energy density $\int_{k_1}^{k_2} dk \, \mathcal{W}(k) =
(\delta B)^2 / 8 \pi$ is small compared with the `unperturbed' magnetic
field energy density, $\zeta \equiv (\delta B)^2 / B_0^2 < 1$. The momentum
diffusion coefficient in equations (\ref{diff1}-\ref{diff2}) can be then
evaluated \citep[e.g.,][]{mel68,kul69,sch89} as
\begin{equation}
D(p) \approx {\zeta \, \beta_{\rm A}^2 \, p^2 \, c \over r_{\rm g}^{2-q} \,
\lambda_2^{q-1}} \, \propto p^q \, ,
\label{coef}
\end{equation}
\noindent
where $\lambda_2 = 2 \pi / k_1$ is the maximum wavelength of the Alfv\'en modes,
$v_{\rm A} \equiv \beta_{\rm A} \, c$ is the Alfv\'en velocity, and $r_{\rm g} =
p c / e B_0$ is the gyroradius of \emph{ultrarelativistic particles}  of
interest here. Similar formulae can be derived for the case of fast magnetosonic
modes \citep[e.g.,][]{kul71,ach81,sch98}. This allows one to find the
characteristic acceleration timescale due to stochastic particle-wave
interactions, $t_{\rm acc} \equiv p^2 / D(p) \propto p^{2-q}/\beta_A^2$.
Similarly, the escape timescale due to particle diffusion from the system of
spatial scale $L$ can be evaluated as $t_{\rm esc} = L^2 / \kappa_{||} \propto
p^{q-2}$, where the spatial diffusion coefficient $\kappa_{||} = (1/3) \, c \,
\Lambda$ is given by the appropriate particle mean free path, $\Lambda \approx
(1/3) \, \zeta^{-1} \, r_{\rm g} \, (\lambda_2/r_{\rm g})^{q-1} \propto
p^{2-q}$, that can be found from the standard relation $D(p) = (1/3) \,
\beta_{\rm A}^2 \, p^2 \, c / \Lambda$ \citep[for more details see,
e.g.,][]{sch02}.

For convenience we define the dimensionless momentum variable $\chi \equiv p /
p_0$, where $p_0$ is some chosen (e.g., injection) particle momentum. With this,
the (stochastic) acceleration and escape timescales can be written as
\begin{eqnarray}
t_{\rm acc} & = & \tau_{\rm acc} \, \chi^{2-q} \, , \quad {\rm where} \quad
\tau_{\rm acc} \equiv {\lambda_2 \over \zeta \, \beta_{\rm A}^2 \, c} \,
\left({p_0 \, c \over e B_0 \, \lambda_2}\right)^{2-q} \, , \nonumber
\\
t_{\rm esc} & = & \tau_{\rm esc} \, \chi^{q-2} \, , \quad {\rm where} \quad
\tau_{\rm esc} \equiv {9 L^2 \, \zeta \over \lambda_2 \, c} \, \left({p_0 \, c
\over e B_0 \, \lambda_2}\right)^{q-2} \, . \label{timescales}
\end{eqnarray}
\noindent
Hereafter we also consider regular energy changes, strictly energy losses, being
an arbitrary function of the particle energy as given by the appropriate
timescale $t_{\rm loss}=t_{\rm loss}(p)$, namely $\langle \dot{p} \rangle = - (p
/ t_{\rm loss})$. We define further $\tau \equiv t / \tau_{\rm acc}$, $N(\chi,
\tau) \equiv p_0 \, n(p, t) \, V$, and $Q(\chi, \tau) \equiv \tau_{\rm acc} \,
p_0 \, \widetilde{Q}(p, t) \, V$, where $V = \int d^3 x$ is the volume of the
system. With such, the momentum diffusion equation (\ref{diff2}) reads as
\begin{equation}
{\partial N \over \partial \tau} = {\partial \over \partial \chi} \left[\chi^q
\, {\partial N \over \partial \chi} \right] - {\partial \over \partial \chi}
\left[\left(2 \, \chi^{q-1} - \chi\, \vartheta_{\chi}\right) N\right] -
\varepsilon \, \chi^{2-q} \, N + Q \, ,
\label{eqfinal}
\end{equation}
\noindent
or, in its steady-state ($\partial N / \partial \tau=0$) form, as
\begin{equation}
{\partial \over \partial \chi} \left[\chi^q \, {\partial N \over \partial \chi}
\right] - {\partial \over \partial \chi} \left[\left(2 \, \chi^{q-1} - \chi\,
\vartheta_{\chi}\right) N\right] - \varepsilon \, \chi^{2-q} \, N = - Q \, .
\label{steady}
\end{equation}
\noindent
In the above, we have introduced
\begin{equation}
\vartheta_{\chi} \equiv {\tau_{\rm acc} \over t_{\rm loss}(\chi)} \quad {\rm
and} \quad \varepsilon \equiv {\tau_{\rm acc} \over \tau_{\rm esc}} \, .
\label{Gamma-definition}
\end{equation}
\noindent

Some specific solutions to the equation (\ref{eqfinal}) were presented in the
literature. Majority of investigations concentrated on the `hard-sphere
approximation' with $q=2$, i.e. with the mean free path for particle-wave
interaction independent of particle energy ($\Lambda = \zeta \, \lambda_2 / 3$;
`classical' Fermi-II process). It was found, that in the absence of regular
energy losses ($\vartheta_{\chi} = 0$), the steady-state solution of equation
(\ref{steady}) with the source term $Q(\chi) \propto \delta(\chi-\chi_{\rm
inj})$, where $\delta(\chi)$ is the Dirac delta, is  of a power-law form
$N(\chi>\chi_{\rm inj}) \propto \chi^{-\sigma}$ with $\sigma = -(1/2) + [(9/4) +
\varepsilon]^{1/2}$ \citep{dav56,ach79,par95}. Note, that for $\varepsilon \ll
1$ this can be approximated by $\sigma \approx 1 + \varepsilon/3$, which is the
original result obtained by \citet{fer49}. In addition, with the increasing
escape timescale, $\varepsilon \rightarrow 0$, the steady-state solution
approaches $N(\chi>\chi_{\rm inj}) \propto \chi^{-1}$. This agrees with the
general finding that for the range $1 \leq q < 2$ and the same injection
conditions the steady-state particle energy distribution implied by the equation
(\ref{steady}) is $N(\chi>\chi_{\rm inj}) \propto \chi^{1-q}$, as long as the
regular energy changes and particle escape can be neglected
\citep[$\vartheta_{\chi} = \varepsilon = 0$;][]{lac79,bor86,dro86,bec06}. The
whole energy range $0 \leq \chi \leq \infty$ with the appropriate (singular)
boundary conditions is considered in \citet{par95}.

The analytic investigations of the momentum diffusion equation (\ref{eqfinal})
in the $q=2$ limit including the radiative cooling have concentrated on the
synchrotron-type losses $\vartheta_{\chi} \propto \chi$ \citep[see,
however,][]{sch87,ste88}. The extended discussion on the time-dependent
evolution for such a case (equation \ref{eqfinal}) was presented by
\citet{kar62}. As for the steady-state solution (equation \ref{steady}), it was
found that with $Q(\chi) \propto \delta(\chi-\chi_{\rm inj})$ and the range $0
\leq \chi \leq \infty$ 
\begin{equation}
N(\chi>\chi_{\rm inj}) \propto \chi^{\sigma+1} \, e^{- {\chi \over \chi_{\rm
eq}}} \, U\!\left[\sigma - 1, \, 2 \, \sigma +2, \, {\chi \over \chi_{\rm
eq}}\right] \, ,
\label{park}
\end{equation}
\noindent
where $\sigma$ is the energy spectral
index introduced above, the equilibrium momentum $\chi_{\rm eq}$ is defined by
the $t_{\rm acc} = t_{\rm loss}$ condition (yielding $\vartheta_{\chi} = \chi /
\chi_{\rm eq}$), and $U[a,b,z]$ is a Tricomi confluent hypergeometrical function
\citep{jon70,sch84,par95}\footnote{The effects of regular energy gains were
omitted here for clarity. \citet{jon70,sch84} and \citet{par95} included in
their investigations regular energy gains representing very idealized shock
acceleration process. Within the anticipated `hard-sphere' approximation, these
gains were assumed to be characterized by the appropriate timescale independent
on the particle energy, $\langle \dot{p}\rangle_{\rm gain} \propto p$ .}. For
$\chi \ll \chi_{\rm eq}$, i.e. for the particle momenta low enough to neglect
radiative losses, the above distribution function has, as expected, a power-law
form $N(\chi > \chi_{\rm inj}) \propto p^{ - \sigma}$. For $\chi \gtrsim
\chi_{\rm eq}$ and $\varepsilon \ll 1$, the particle energy spectrum approaches
$N(\chi > \chi_{\rm inj}) \propto \chi^2 \, \exp \left( - \chi/\chi_{\rm
eq}\right)$. That is, as long as particle escape is inefficient, a two component
stationary energy distribution is formed: a power-law $\propto \chi^{-1}$ at low
($\chi < \chi_{\rm eq}$) energies, and a pile-up bump (`ultrarelativistic
Maxwellian distribution') around $\chi \sim \chi_{\rm eq}$. For  shorter
escape timescale no pile-up form appears, and the resulting particle spectral
index depends on the ratio $\varepsilon$ of the escape and the acceleration time
scales.

In the case of $q \neq 2$ and synchrotron-type energy losses
$\vartheta_{\chi} \propto \chi$, the steady-state solution to the equation
(\ref{steady}) provided by \citet{mel69} was questioned due to unclear boundary
conditions applied \citep{tad71,par95}. The special case of $q = 1$ with
particle escape included (and the infinite energy range $0 \leq \chi \leq
\infty$) was considered further by \citet{bog85}. It was found, that with the
injection of the $Q(\chi) \propto \delta(\chi-\chi_{\rm inj})$ type, the steady
state solution of equation (\ref{steady}) is\footnote{The other
(non-synchrotron) radiative losses terms included in the analysis presented by
\citet{bog85} were omitted here for clarity.}
\begin{equation}
N(\chi>\chi_{\rm inj}) \propto \chi^{2} \, e^{- {1\over 2} \, \left({\chi \over
\chi_{\rm eq}}\right)^2} \, U\!\left[{1 \over 2} \left({\chi_{\rm eq} \over
\chi_{\rm esc}}\right)^2 , \, 2 \, , \, {1 \over 2} \left({\chi \over \chi_{\rm
eq}}\right)^2\right] \, ,
\label{bogdan}
\end{equation}
\noindent
where the critical escape and equilibrium momenta $\chi_{\rm esc}$ and
$\chi_{\rm eq}$ are defined by the conditions $t_{\rm esc} = t_{\rm acc}$ and
$t_{\rm acc} = t_{\rm loss}$, respectively, yielding $\varepsilon = 1/ \chi_{\rm
esc}^{2}$ and $\vartheta_{\chi} = \chi / \chi_{\rm eq}^2$. This solution implies
$N(\chi > \chi_{\rm inj}) \propto const$ at low particle momenta for which
synchrotron energy losses are negligible ($\chi \ll \chi_{\rm eq}$),
independent of the particular value of the escape timescale. At higher
particle energies, an exponential dependence is expected, $N(\chi > \chi_{\rm
inj}) \propto \chi^{2 - (\chi_{\rm eq}/\chi_{\rm esc})^2} \, \exp \left[ - {1
\over 2} \, \left(\chi/\chi_{\rm eq}\right)^2 \right]$. Note, that with an
increasing escape timescale this approaches
$\sim \chi^2 \, \exp \left[ - {1 \over 2} \, \left(\chi/\chi_{\rm eq}\right)^2
\right]$.

\section{Inefficient Particle Escape}

In this section we are interested in steady-state solutions to the momentum
diffusion equation (\ref{steady}) in the case of a very inefficient particle
escape and a general (i.e., not necessarily synchrotron-type) form of the
regular energy changes $\vartheta_{\chi}$, which is however a continuous
function of the particle energy. With $\varepsilon = 0$, the homogeneous form of
this equation can be therefore transformed to the self-adjoint form
\begin{equation}
{d \over d \chi} \left[ P(\chi) \, {d \over d \chi} N(\chi)\right] - G(\chi) \,
N(\chi) = 0
\label{self}
\end{equation}
\noindent
with
\begin{eqnarray}
P(\chi)  & = & \chi^q \, S(\chi) \, , \nonumber \\
G(\chi)  & = & \left[ 2 (q-1) \chi^{q-2} - {d \over d \chi}\left(\chi \,
\vartheta_{\chi}\right)\right] \, S(\chi) \, , \nonumber\\
S(\chi) & = & \chi^{-2} \, \exp\left[\int^\chi d \chi' \, \chi'^{1-q} \,
\vartheta_{\chi'}\right] \, . \label{s}
\end{eqnarray}
\noindent
We also restrict the analysis to the finite particle energy range $\chi \in
[\chi_1, \, \chi_2]$, where $0 < \chi_1, \, \chi_2 < \infty$. The justification
for this is that for a finite range of the turbulent wavevectors, say $k
\in [k_1, \, k_2]$, the momentum diffusion coefficient as given in
equation (\ref{coef}) is well defined only for a finite range of particle
energies (momenta). For example, gyroresonant interactions between the particles
and the Alf\'enic turbulence require particles' gyroradii comparable to the
scale of the interacting modes, or $k\,r_{\rm g} \sim 2\pi$. Hence, the lower
and upper limit of the particle energy range could be chosen as $\chi_1 =
2\pi e B_0 / k_1 c p_0$ and $\chi_2 = 2\pi e B_0 /k_2 c p_0$,
respectively\footnote{In the case of the magnetosonic-type turbulence,
interacting with particles via transit-time damping satisfying the Cherenkov
condition $k\,r_{\rm g} \ll 1$, the low energy cut-off in the momentum
diffusion coefficient could be chosen to be the energy of the particle whose
velocity is comparable to the velocity of the fast magnetosonic mode, which is
$\sim v_A$ for low-$\beta$, or magnetically dominated plasmas.}. Since all of the
functions $P(\chi)$, $P'(\chi)$, $G(\chi)$, $S(\chi)$ are continuous, and
$P(\chi)$, $S(\chi)$ are finite and strictly positive in the considered (closed)
energy interval, the appropriate boundary value problem,
\begin{eqnarray}
a_1 \, N(\chi_1) + a_2 \, \left. {d N(\chi) \over d \chi}\right|_{\chi_1} & = & 0 \,
, \nonumber \\
b_1 \, N(\chi_2) + b_2 \, \left. {d N(\chi) \over d \chi}\right|_{\chi_2} & = & 0 \,
, \label{bc}
\end{eqnarray}
\noindent
is regular. If one of these conditions is violated, which is the case for the
infinite energy range $0 \leq \chi \leq \infty$, the problem becomes singular,
and the extended analysis presented by \citet{par95} has to be applied.

The two linearly-independent particular solutions to the homogeneous form of the
equation (\ref{self}) are
\begin{eqnarray}
y_1(\chi) & = & S^{-1}(\chi) \, , \nonumber \\
y_2(\chi) & = & S^{-1}(\chi) \, \int^\chi d\chi' \, \chi'^{-q} \, S(\chi') \, ,
\label{y}
\end{eqnarray}
\noindent
or any linear combination of these,
\begin{eqnarray}
u_1(\chi) & = & y_1(\chi) + \alpha \, y_2(\chi) \, , \nonumber \\
u_2(\chi) & = & y_1(\chi) + \beta \, y_2(\chi) \label{u}
\end{eqnarray}
\noindent
(each involving arbitrary multiplicative constants). By imposing the boundary
conditions (\ref{bc}) in a form
\begin{eqnarray}
a_1 \, u_1(\chi_1) + a_2 \, \left .{d u_1(\chi) \over d \chi}\right|_{\chi_1} & = &
0 \, , \nonumber \\
b_1 \, u_2(\chi_2) + b_2 \, \left. {d u_2(\chi) \over d \chi}\right|_{\chi_2} & = &
0 \, , \label{bvp}
\end{eqnarray}
\noindent
parameters $\alpha$ and $\beta$ can be determined. With thus constructed
particular solutions to the equation (\ref{self}), one can define the Wronskian
$w(\chi) \equiv u_1(\chi) \, u_2'(\chi) - u_1'(\chi) \, u_2(\chi)$, and next
construct the Green's function of the problem,
\begin{equation}
\mathcal{G}(\chi, \chi_{\rm inj}) = {1 \over - \chi_{\rm inj}^q \, w(\chi_{\rm inj})} \times \left\{
\begin{array}{ccc} 
u_1(\chi) \, u_2(\chi_{\rm inj}) & {\rm for} & \chi_1 \leq \chi < \chi_{\rm inj} \\
u_1(\chi_{\rm inj}) \, u_2(\chi)  & {\rm for} & \chi_{\rm inj} < \chi \leq \chi_2
\end{array} \right. \, ,
\label{green1}
\end{equation}
\noindent
where $\chi_1 < \chi_{\rm inj} < \chi_2$. This gives the final solution to the equation
(\ref{steady})
\begin{equation}
N(\chi) = \int_{\chi_1}^{\chi_2} d\chi_{\rm inj} \, \mathcal{G}(\chi, \chi_{\rm inj}) \, Q(\chi_{\rm inj}) \, .
\label{solution}
\end{equation}
\noindent

Steady-state solutions exist, however, only for some particular forms of the
injection function $Q(\chi,\tau)$. To investigate this issue, and to impose
correct boundary conditions for the finite energy range $\chi_1 \leq \chi \leq
\chi_2$, let us integrate equation (\ref{eqfinal}) over the
energies and re-write it in a form of the continuity equation,
\begin{equation}
{\partial \mathcal{N} \over \partial \tau} + \left. \mathcal{F}\right|_{\chi_2}
- \left. \mathcal{F}\right|_{\chi_1} = \int_{\chi_1}^{\chi_2} d\chi \, Q(\chi,
\tau) \, .
\label{cont}
\end{equation}
\noindent
Here $\mathcal{N} \equiv \int_{\chi_1}^{\chi_2} d\chi \, N(\chi)$ is the total
number of particles and the particle flux in the momentum space is defined as
\begin{equation}
\mathcal{F}[N(\chi)] = \left(2 \chi^{q-1} - \chi \, \vartheta_{\chi} \right) N -
\chi^q \, {\partial N \over \partial \chi} \, .
\label{flux1}
\end{equation}
\noindent
Note, that with the particular solutions $u_1(\chi)$ and $u_2(\chi)$ given in
(\ref{u}), one has
\begin{eqnarray}
\mathcal{F}[u_1(\chi)] & = & - \alpha \, , \nonumber \\
\mathcal{F}[u_2(\chi)] & = & - \beta \, , \label{flux2}
\end{eqnarray}
\noindent
independent of the momentum $\chi$ or of the particular form of the direct
energy losses function $\vartheta_{\chi}$.

Let us comment in this context on the `zero-flux' boundary conditions of the
type (\ref{bc}), namely $\left.\mathcal{F}\right|_{\chi_1} =
\left.\mathcal{F}\right|_{\chi_2} = 0$. These, with equations (\ref{bvp}) and
(\ref{flux2}), imply $\alpha = \beta = 0$, i.e., $u_1(\chi) = u_2(\chi)$. In
other words, one particular solution $y_1(\chi)$ satisfies the `no-flux'
boundary condition of the homogeneous form of the equation (\ref{self}) for both
$\chi_1$ and $\chi_2$. In such a case, the steady-state solution can be
constructed using the function $y_1(\chi)$ only if it is orthogonal to the
source function, $\int_{\chi_1}^{\chi_2} d\chi \, y_1(\chi) \, Q(\chi) = 0$.
This condition, for any non-zero particle injection and $y_1(\chi) =
S^{-1}(\chi)$ as given in the equation (\ref{s}), cannot be fulfilled
\citep[cf.][]{mel69,tad71}. `Zero-flux' boundary conditions for non-vanishing
$Q(\chi)$ can be instead imposed if the particle injection is balanced by the
particle escape from the system (see \S\,3 below).

In the case of no particle escape, with the stationary injection such that
$\int_{\chi_1}^{\chi_2} d\chi \, Q(\chi) \equiv A$ and with the direct
(radiative) energy losses $\vartheta_{\chi} \neq 0$, the boundary conditions can
be chosen as
\begin{equation}
- \left.\mathcal{F}\right|_{\chi_1} = A \, , \quad {\rm and} \quad
\left.\mathcal{F}\right|_{\chi_2} = 0 \, ,
\label{conditions} 
\end{equation}
\noindent
which give $\alpha = A$ and $\beta = 0$, and correspond to the conservation of
the total number of particles within the energy range $[\chi_1, \chi_2]$. Let us
justify this choice by noting that the radiative losses processes, unlike
momentum diffusion strictly related to the turbulence spectrum, is well defined
for particle momenta $\chi<\chi_1$ and $\chi> \chi_2$. Hence, with non-vanishing
radiative losses (as expected for ultrarelativistic particles considered here),
no flux of particles in the momentum space through the maximum value $\chi_2$
toward higher energies is possible (radiative losses in the absence of
stochastic acceleration will always prevent from presence of particle flux above
$\chi_2$). For the same reason, there is always a possibility for a non-zero
particle flux toward lower energies through the $\chi_1$ point, since the
stochastic acceleration timescale, even if being an increasing function of the
particle energy, is always finite at $\chi_1 > 0$. Note in this context, that
the particle flux at $\chi_1$ implied by the chosen boundary conditions
(\ref{conditions}) must be negative, $\left.\mathcal{F}\right|_{\chi_1} < 0$.
That is, there is a continuous flux of particles through the $\chi_1$ point from
high to low energies, which --- in the absence of particle catastrophic escape
from the system --- balances particle injection. With these, one can find the
Green's function as
\begin{equation}
\left. \mathcal{G}(\chi, \chi_{\rm inj})\right|_{\rm loss} = \left\{ \begin{array}{ccc}
S^{-1}(\chi) \, \left(A^{-1} + \int^\chi_{\chi_1} d\chi' \, \chi'^{-q} \,
S(\chi') \right) & {\rm for} & \chi_1 \leq \chi < \chi_{\rm inj} \\
S^{-1}(\chi) \, \left(A^{-1} + \int^{\chi_{\rm inj}}_{\chi_1} d\chi' \, \chi'^{-q} \,
S(\chi') \right) & {\rm for} & \chi_{\rm inj} < \chi \leq \chi_2
\end{array} \right. \, ,
\label{green-loss}
\end{equation}
\noindent
where $S(\chi)$, introduced in the equation (\ref{s}), can be re-written as
\begin{equation}
S(\chi) = \chi^{-2} \, \exp\left[\int^\chi {d \chi' \over \chi'} \, {t_{\rm
acc}(\chi') \over t_{\rm loss}(\chi')}\right] \, . 
\label{s-loss}
\end{equation}
\noindent

\subsection{Synchrotron Energy Losses}

As an example let us consider synchrotron energy losses of ultrarelativistic
electrons, which are characterized by the timescale 
\begin{equation}
t_{\rm syn} = \tau_{\rm syn} \, \chi^{-1} \, , \quad {\rm with} \quad \tau_{\rm
syn} \equiv {6 \pi \, m_{\rm e}^2 c^2 \over \sigma_{\rm T} \, p_0 \, B_0^2}
\label{syn}
\end{equation}
\noindent
\citep[e.g.,][]{blu70}, and which define the equilibrium momentum $\chi_{\rm eq}
= (\tau_{\rm syn} / \tau_{\rm acc})^{1/(3-q)}$ through the condition $t_{\rm
acc} = t_{\rm syn}$, yelding $\vartheta = \chi / \chi_{\rm eq}^{3-q}$. The
Green's function (\ref{green-loss}) adopts then the form
\begin{eqnarray}
& & \left. \mathcal{G}(\chi, \chi_{\rm inj})\right|_{\rm syn} = \chi^2 \, e^{- \, {1 \over 3-q}
\, \left({\chi \over \chi_{\rm eq}}\right)^{3-q}} \, \left({1 \over A} +
\int^{\min[\chi_{\rm inj},\, \chi]}_{\chi_1} d\chi' \, \chi'^{-(2+q)} \, e^{{1 \over 3-q}
\, \left({\chi' \over \chi_{\rm eq}}\right)^{3-q}} \right) =  \label{green-syn}
\\
& & = \chi^2 \, e^{- \, {1 \over 3-q} \, \left({\chi \over \chi_{\rm
eq}}\right)^{3-q}} \, \left( {1 \over A} + {\chi_{\rm eq}^{-1-q} \, (-1)^{4 /
(3-q)} \over (3-q)^{4 / (3-q)}} \, \Gamma\left[ - {1+q \over 3-q} \, , \, - \,
{\left(\min[\chi_{\rm inj},\, \chi] / \chi_{\rm eq}\right)^{3-q} \over 3-q} , \, - \,
{\left(\chi_1 / \chi_{\rm eq}\right)^{3-q} \over 3-q} \right] \right) \, ,
 \nonumber
\end{eqnarray}
\noindent
where $\Gamma[a, z_1, z_2]$ is generalized incomplete Gamma function. By
expressing the above solution in terms of Kummer confluent hypergeometrical
functions $M[a,b,z]$ using the identity $\Gamma[a,z_1,z_2] = a^{-1} \, z_2^a \,
M[a, 1+a, -z_2] - a^{-1} \, z_1^a \, M[a, 1+a, -z_1]$ \citep{abr64}, assuming
$\chi_1 \ll \chi_{\rm eq}$, and noting that $M[a,b,z] \sim 1$ for $z \rightarrow
0$, one can rewrite it further as
\begin{eqnarray}
& & \left. \mathcal{G}(\chi, \chi_{\rm inj})\right|_{\rm syn} \approx \chi^2 \, e^{- \, {1
\over 3-q} \, \left({\chi \over \chi_{\rm eq}}\right)^{3-q}} \times
\label{G-syn-app1} \\
& & \times \left({1 \over A} + {\chi_1^{-1-q} \over 1+q} - {\min(\chi_{\rm inj},\,\chi)^{-1-q}
\over 1+q} \, M\left[-{1+q \over 3-q},\, {2-2q \over 3-q},\, {1 \over 3-q}
\left({\min[\chi_{\rm inj},\, \chi] \over \chi_{\rm eq}}\right)^{3-q}\right]\right) \, .
\nonumber
\end{eqnarray}
\noindent
Finally, noting that $M[a,b,z] \sim \Gamma(b) \, e^z \, z^{a-b} / \Gamma(a)$ for
$z \rightarrow \infty$, and neglecting the $A^{-1}$ term, one finds a rough
approximation 
\begin{equation}
\left. \mathcal{G}(\chi, \chi_{\rm inj})\right|_{\rm syn} \sim 
\left\{ \begin{array}{ccc}
{1 \over 1+q} \, \chi_1^{-1-q} \, \chi^2 \, e^{- \, {1 \over 3-q} \, \left(\chi
/ \chi_{\rm eq}\right)^{3-q}} \quad & {\rm for} & \min(\chi_{\rm inj}, \chi) \lesssim
\chi_{\rm eq} \\
\chi_{\rm eq}^{3-q} \, \chi^{-2} \quad & {\rm for} & \chi_{\rm eq} \ll \chi <
\chi_{\rm inj} \\
\chi_{\rm eq}^{3-q} \, \chi_{\rm inj}^{-4} \, e^{{1 \over 3-q} \, \left(\chi_{\rm inj} / \chi_{\rm
eq}\right)^{3-q}} \, \chi^2 \, e^{- \, {1 \over 3-q} \, \left(\chi / \chi_{\rm
eq}\right)^{3-q}} \quad & {\rm for} & \chi_{\rm eq} \ll \chi_{\rm inj} < \chi \end{array}
\right. \, .
\label{G-syn-app2}
\end{equation}
\noindent
Hence, as long as $\min[\chi_{\rm inj}, \, \chi] < \chi_{\rm eq}$, one has $\left.
\mathcal{G}(\chi, \chi_{\rm inj})\right|_{\rm syn} \propto \chi^2 \, \exp\left[- {1\over
3-q} \left(\chi / \chi_{\rm eq}\right)^{3-q}\right]$. If, however, $\min[\chi_{\rm inj}, \,
\chi] > \chi_{\rm eq}$, the Green's function retains the spectral shape $\propto
\chi^2 \, \exp\left[- {1\over 3-q} \left(\chi / \chi_{\rm
eq}\right)^{3-q}\right]$ for $\chi_{\rm inj} < \chi$, while is of a power-law form $\left.
\mathcal{G}(\chi, \chi_{\rm inj})\right|_{\rm syn} \propto \chi^{-2}$ for $\chi < \chi_{\rm inj}$. 

\begin{figure}
\centering
\includegraphics[scale=1.6]{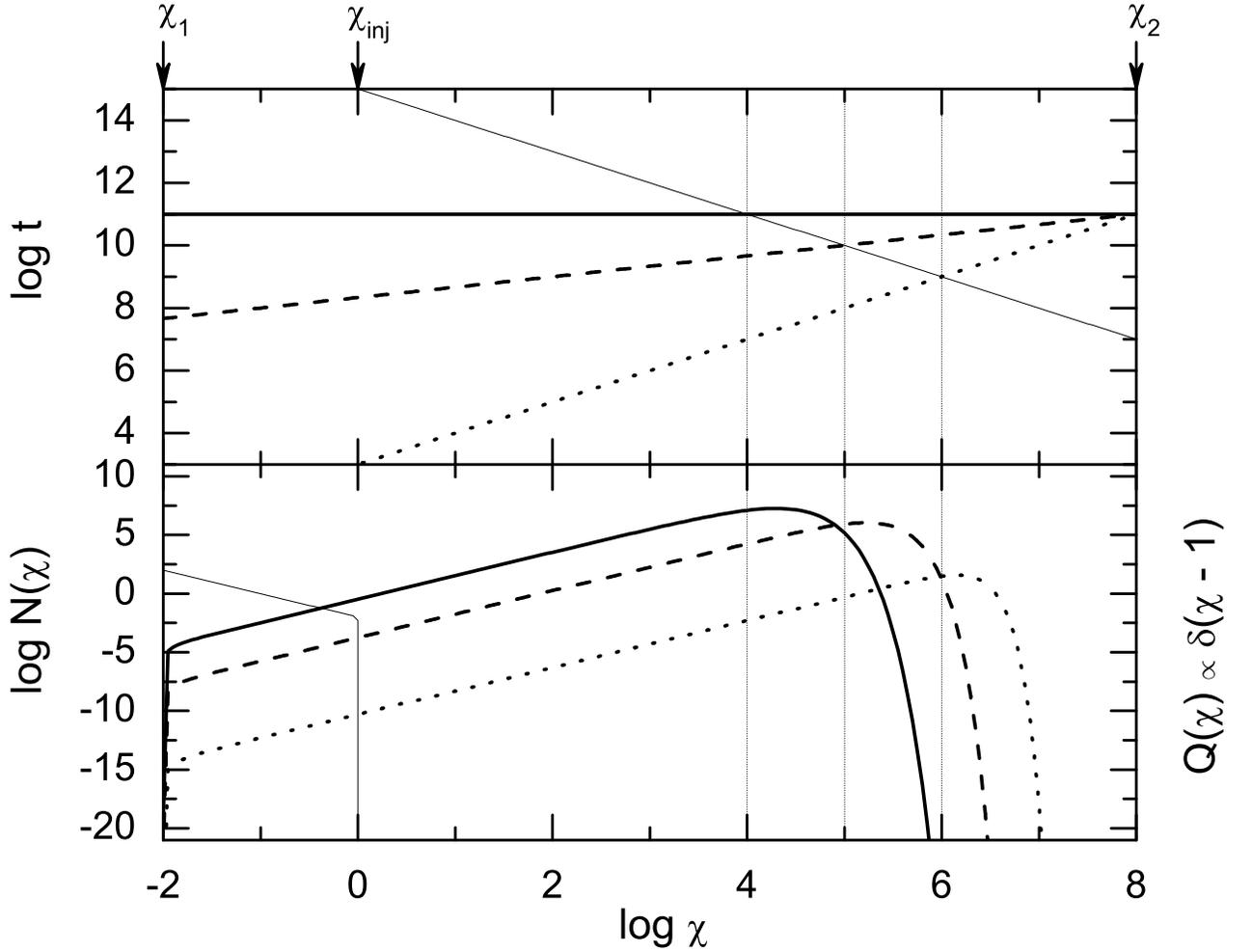}
\caption{{\it Upper panel:} Stochastic acceleration timescales for fixed plasma
parameters ($B_0$, $\zeta$, $\beta_{\rm A}$, $\chi_1$, $\chi_2$) but different
turbulence energy index: $q=2$ (thick solid lines), $q = 5/3$ (thick dashed
lines), and $q=1$ (thick dotted lines). Thin solid line denotes radiative
(synchrotron) energy losses timescale considered. {\it Lower panel:} Particle
spectra resulting from joint stochastic acceleration and radiative (synchrotron)
energy losses specified in the upper panel. The spectra correspond to the
monoenergetic injection $Q(\chi) \propto \delta(\chi-1)$ with fixed $\int dp
\, \widetilde{Q}(p)$, and no particle escape. Thin solid line denotes particle
spectrum expected for the same injection and cooling conditions, but with the
momentum diffusion effects neglected, $\widetilde{N}(\chi)$.}
\label{delta}
\end{figure}

\begin{figure}
\centering
\includegraphics[scale=1.6]{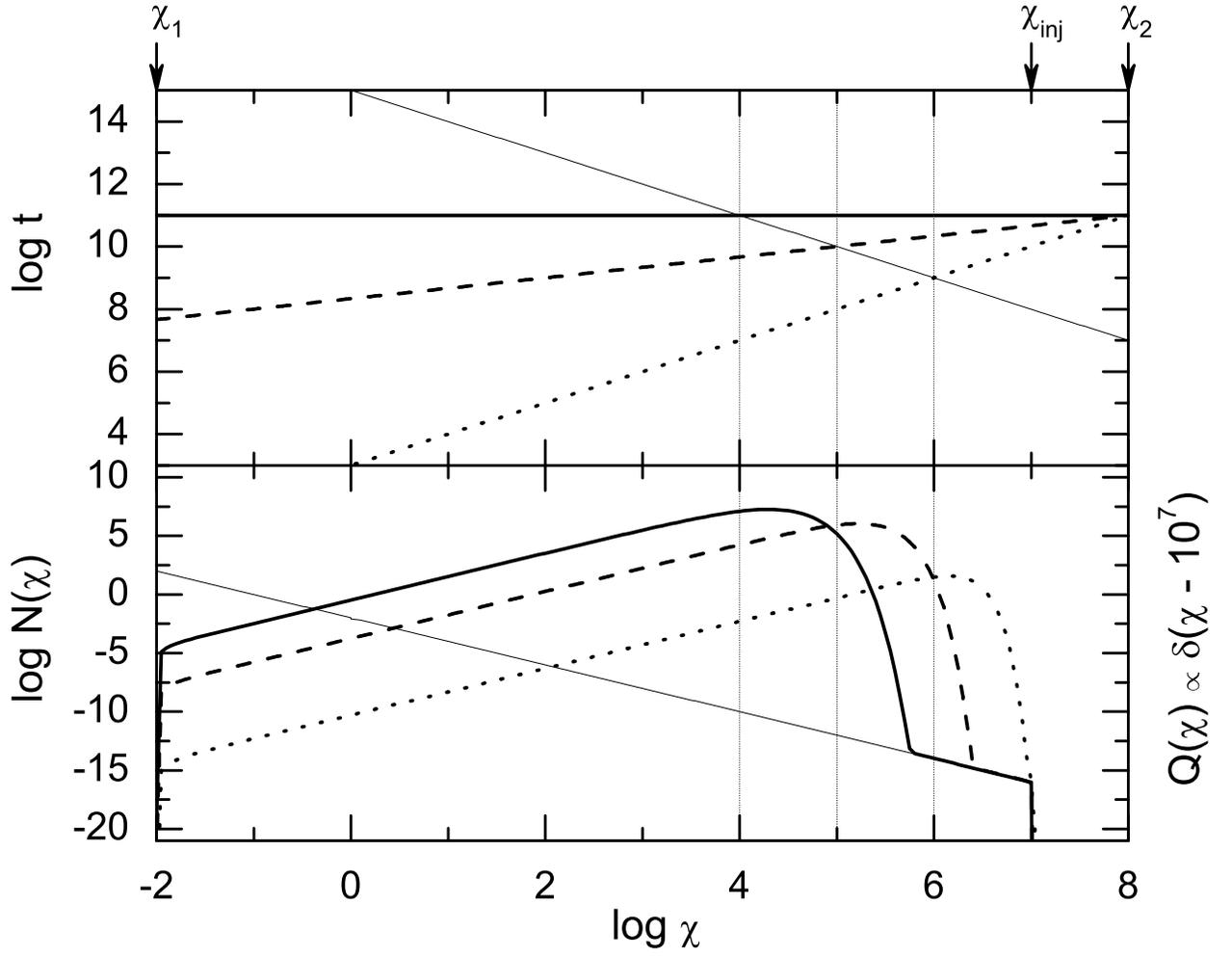}
\caption{The same as Figure (\ref{delta}) except for $Q(\chi) \propto \delta(\chi-10^7)$.}
\label{powerlaw}
\end{figure}

In Figures (\ref{delta}) and (\ref{powerlaw}) we plot examples of particle
spectra obtained from the above solution for the system with fixed plasma
parameters ($B_0$, $\zeta$, $\beta_{\rm A}$, $\chi_1$, $\chi_2$) but different
turbulence energy indices: $q=2$ (`hard-sphere' approximation; thick solid lines
in the figures), $q = 5/3$ (Kolmogorov-type turbulence; thick dashed lines), and
$q=1$ (Bohm limit; thick dotted lines). As for the source function, we consider
two different forms, namely $Q(\chi) \propto \delta(\chi-1)$ in Figure
\ref{delta} and $Q(\chi) \propto \delta(\chi-10^7)$ in Figure \ref{powerlaw},
with the normalizations given in both cases by the same fixed $\int dp \,
\widetilde{Q}(p)$. The emerging particle spectra are compared with the ones
expected for the same injection and cooling conditions, but with the momentum
diffusion neglected, $\widetilde{N}(\chi)$. Such a steady-state electron
distribution can be found from the appropriate equation
\begin{equation}
{\partial \over \partial \chi} \left[\chi\, \vartheta_{\chi} \,
\widetilde{N}(\chi)\right] + Q(\chi) = 0
\label{kinetic}
\end{equation}
\noindent
(see equation \ref{eqfinal}), for which one has the straightforward solution
\citep{kar62}
\begin{equation}
\widetilde{N}(\chi) = {1 \over \chi \, \vartheta_{\chi}} \, \int_{\chi}^{\chi_2}
Q(\chi_{\rm inj}) \, d\chi_{\rm inj} 
\label{kardashev}
\end{equation}
\noindent
(thin solid lines in the lower panels of Figures \ref{delta}-\ref{powerlaw}). 

As shown in the figures and follows directly from the obtained solution
\ref{green-syn}--\ref{G-syn-app2}, joint stochastic acceleration and radiative
(synchrotron-type) loss processes, in the absence of particle escape, tend to
establish $N(\chi) \propto \chi^2 \, \exp\left[- {1 \over 3-q} \, \left(\chi /
\chi_{\rm eq}\right)^{3-q}\right]$ spectra independent of the energy of the
injected particles and the form of the source function as long as it has a
narrow distribution. Moreover, for $\chi \ll \chi_{\rm eq}$ the turbulence
energy index $q$ does not influence the spectral shape of the electron energy
distribution. Instead --- with fixed normalization of the injection function
$\widetilde{Q}(p)$ and fixed plasma parameters (including magnetic field
intensities $B_0$ and $\zeta$) --- turbulence power-law slope $q$ determines (i)
the equilibrium momentum $\chi_{\rm eq}$, (ii) normalization of the electron
energy distribution, and (iii) the spectral shape of the particle distribution
for $\chi \geq \chi_{\rm eq}$. In particular, flatter turbulent spectrum leads
to higher value of the equilibrium momentum $\chi_{\rm eq}$, lower normalization
of $N(\chi)$, and steeper exponential cut-off at $\chi > \chi_{\rm eq}$. Note
also, that if particles with momenta $\chi_{\rm inj} \gg \chi_{\rm eq}$ are
being injected to the system, the resulting electron energy distribution may
adopt the `standard' form of the synchrotron-cooled source function
(\ref{kardashev}) at highest momenta $\chi_{\rm eq} \ll \chi < \chi_{\rm inj}$
(e.g., $\propto \chi^{-2}$ for the $Q(\chi) \propto \delta(\chi - 10^7)$
injection in Figure \ref{powerlaw}).

\subsection{Inverse-Compton Energy Losses and the Klein-Nishina Effects}

Let us now investigate the effects of the inverse-Compton (IC) radiative energy
losses in the presence of a turbulent particle acceleration. At low energies
when the Klein-Nishina (KN) effects are negligible the IC case is identical to
the synchrotron case with the magnetic energy density $B^2/8\pi$ replaced by the
photon energy density $u_{\rm ph}$. The two cases differ when KN effects become
important at high energies. To include these effects we approximate  the
radiative loss timescale as
\begin{equation}
t_{\rm IC} = \tau_{\rm IC} \, \chi^{-1} \, \left(1 + {\chi \over \chi_{\rm
cr}}\right)^{1.5} \, , \quad {\rm where} \quad \tau_{\rm IC} \equiv {3 \, m_{\rm
e}^2 c^2 \over 4 \, \sigma_{\rm T} \, p_0 \, u_{\rm ph}} \quad {\rm and} \quad
\chi_{\rm cr} \equiv {m_{\rm e} c \over 4 \, p_0 \epsilon_0} \, .
\label{ic}
\end{equation}
\noindent
Here the radiation field involved in the IC scattering was assumed to be
monoenergetic, with the total energy density $u_{\rm ph}$ and the dimensionless (i.e.,
expressed in the electron mass units) photon energy $\epsilon_0$. The above
formula properly takes into account KN effect up to energies $\chi \leq 10^4 \,
\chi_{\rm cr}$ \citep{mod05}. Clearly, as long as $q < 1.5$, balance between
acceleration and cooling timescales takes place at one particular momentum
$\chi_{\rm eq}= \max(\chi_{\rm Th}, \, \chi_{\rm KN})$, depending on weather
energy losses dominate over acceleration in the Thomson regime, $\chi_{\rm eq}=
\chi_{\rm Th} \equiv (\tau_{\rm IC}/\tau_{\rm acc})^{1/(3-q)}$, or in the KN
regime, $\chi_{\rm eq}= \chi_{\rm KN} \equiv \chi_{\rm Th}^{(3-q)/(1.5-q)} \,
\chi_{\rm cr}^{-1.5 / (1.5 - q)}$. For $q > 1.5$, there may be instead two
equilibrium momenta for a given one acceleration timescale, $\chi_{\rm eq, \, 1}
= \chi_{\rm Th}$ and $\chi_{\rm eq, \, 2} =\chi_{\rm KN}$, or no equilibrium
momentum at all, if $t_{\rm acc} < t_{\rm ic}$ within the whole considered range
$\chi < 10^4 \, \chi_{\rm cr}$. Finally, for $q = 1.5$ (that corresponds to the
Kraichnan turbulence), the ratio between IC/KN and acceleration timescales is
energy-independent, since both $t_{\rm acc} \propto  \chi^{q-1}=\chi^{1/2}$ and,  
as given in (\ref{ic}), $t_{\rm IC}(\chi > \chi_{\rm cr}) \propto \chi^{1/2}$. 

Assuming hereafter $q \neq 3/2$, one can find from the equation (\ref{s-loss})
that
\begin{equation}
S(\chi) = \chi^{-2} \exp\left\{ {1 \over 3-q} \left({\chi \over \chi_{\rm
T}}\right)^{3-q} \, F\!\left[{3 \over 2}, \, 3 - q,\, 4-q,\, -{\chi \over
\chi_{\rm cr}}\right]\right\} \, ,
\label{Gauss}
\end{equation}
where $F[a,b,c,z]$ is Gauss hypergeometric function. This gives the Green's
function
\begin{eqnarray}
& & \left. \mathcal{G}(\chi, \chi_{\rm inj})\right|_{\rm ic}^{q \neq 1.5} = \chi^2 \,
\exp\left\{ - {1 \over 3-q} \left({\chi \over \chi_{\rm Th}}\right)^{3-q} \,
F\!\left[{3 \over 2}, \, 3 - q,\, 4-q,\, -{\chi \over \chi_{\rm
cr}}\right]\right\} \times \label{green-ic} \\
& & \times \left( {1 \over A} + \int_{\chi_1}^{\min(\chi_{\rm inj}, \chi)} d\chi' \,
\chi'^{-2-q} \, \exp\left\{ {1 \over 3-q} \left({\chi' \over \chi_{\rm
Th}}\right)^{3-q} \, F\!\left[{3 \over 2}, \, 3 - q,\, 4-q,\, -{\chi' \over
\chi_{\rm cr}}\right]\right\} \right) \, . \nonumber
\end{eqnarray}
\noindent
Below we discuss some properties of the obtained solution by expanding the Gauss
hypergeometric functions as $F[a,b,c,z] \sim 1$ for $z \rightarrow 0$, and
$F[a,b,c,z] \sim \left[\Gamma(c) \, \Gamma(b-a)/\Gamma(b) \, \Gamma(c-a)\right]
\, (-z)^{-a} + \left[\Gamma(c) \, \Gamma(a-b)/\Gamma(a) \, \Gamma(c-b)\right] \,
(-z)^{-b}$ for $z \rightarrow \infty$.

\begin{figure}
\centering
\includegraphics[scale=1.6]{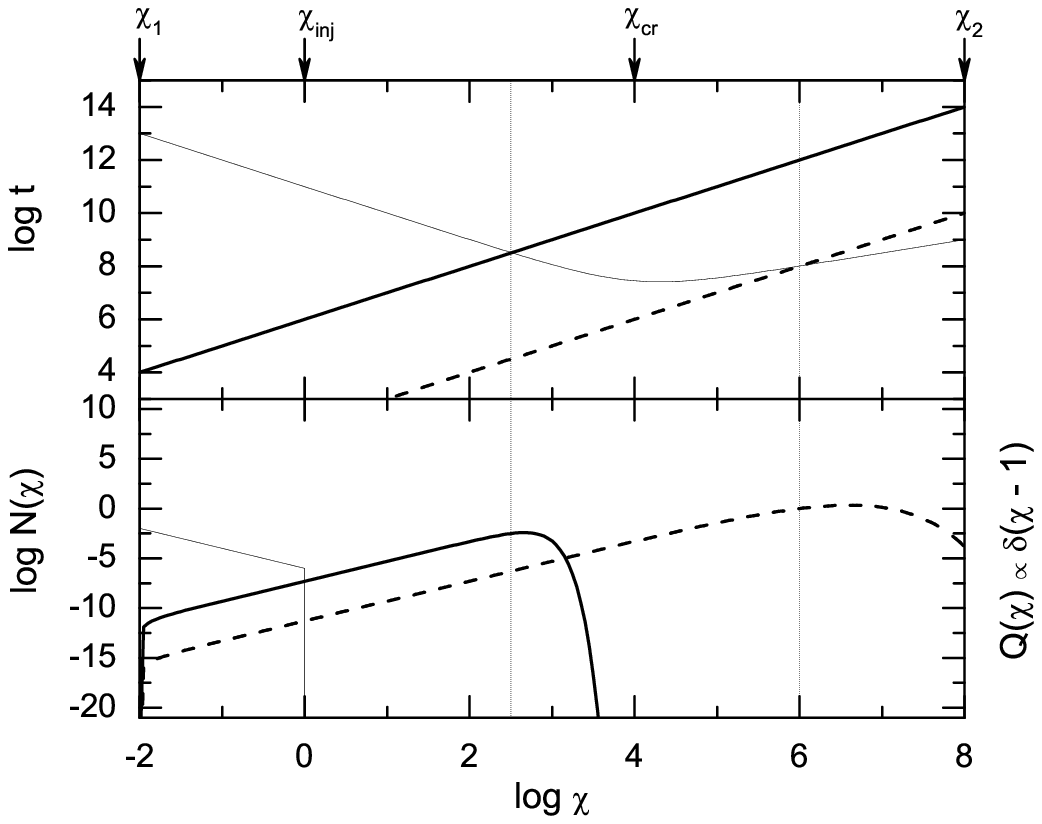}
\caption{{\it Upper panel:} Stochastic acceleration timescales for fixed $q=1$
and different plasma parameters (thick solid and dashed lines). Thin solid line
denotes inverse-Compton energy losses timescale considered with the assumed
$\chi_{\rm cr} = 10^4$. {\it Lower panel:} Particle spectra resulting from joint
stochastic acceleration and inverse-Compton energy losses specified in the upper
panel. The spectra correspond to the monoenergetic injection $Q(\chi) \propto
\delta(\chi-1)$ with fixed $\int dp \, \widetilde{Q}(p)$, and no particle
escape. Thin solid line denotes particle spectrum expected for the same
injection and cooling conditions, but with the momentum diffusion effects
neglected, $\widetilde{N}(\chi)$.}
\label{KN-1}
\end{figure}

\begin{figure}
\centering
\includegraphics[scale=1.6]{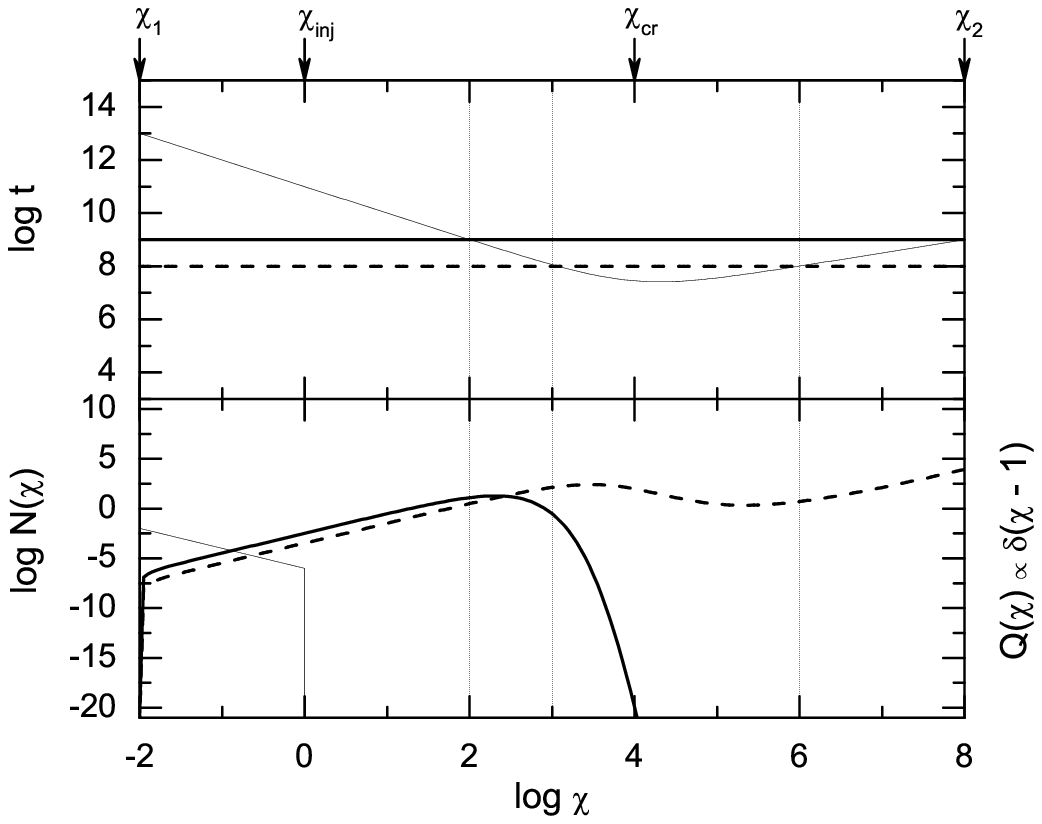}
\caption{The same as Figure (\ref{KN-1}) except for $q=2$.}
\label{KN-2}
\end{figure}

Let us consider first the case of a low-energy injection, such that $\chi_{\rm inj} <
\min(\chi_{\rm Th}, \, \chi_{\rm cr})$. The Green's function (\ref{green-ic}) can
be then approximated roughly by
\begin{equation}
\left. \mathcal{G}(\chi, \chi_{\rm inj})\right|_{\rm ic; \, \chi_{\rm inj}<}^{q \neq 1.5} \sim 
\left\{ \begin{array}{ccc}
{1 \over 1+q} \, \chi_1^{-1-q} \, \chi^2 \, e^{- \, {1 \over 3-q} \, \left(\chi
/ \chi_{\rm Th}\right)^{3-q}} \quad & {\rm for} & \chi < \chi_{\rm cr} \\
{1 \over 1+q} \, \chi_1^{-1-q} \,  
e^{- \, {2 \over \sqrt{\pi}} \, \Gamma(3-q) \, \Gamma(q-1.5) \, \left(\chi_{\rm
cr} / \chi_{\rm Th}\right)^{3-q}} \, \chi^2 \, e^{- \, {1 \over 1.5-q} \,
\left(\chi / \chi_{\rm KN}\right)^{1.5-q}} \,\quad & {\rm for} & \chi > \chi_{\rm cr}
\end{array} \right. \, .
\label{G-ic-app1}
\end{equation}
\noindent
For $\chi < \chi_{\rm cr}$ the Green's function has the same form as  the
synchrotron case, which is expected in the Thomson regime. However, $\chi >
\chi_{\rm cr}$, the  KN effects
modify the high-energy segment of the particle energy distribution. In
particular, for $q < 1.5$ (e.g. $q=1$ in Fig. \ref{KN-1}) the spectrum is always of a `single-bump' form,
possessing  either a sharp or a smooth exponential cut-off depending on whether
we are in the Thomson or KN cooling regime, respectively. On the other hand, with $q > 1.5$ (e.g. $q=2$ in Fig. \ref{KN-2}) 
the acceleration and loss timescales can be equal at two different energies, 
in which case the particle spectra become concave, flattening smoothly from the exponential decrease $\propto
\chi^2 \, \exp\left[- {1\over 3-q} \left(\chi / \chi_{\rm
Th}\right)^{3-q}\right]$ at $\chi_{\rm Th} < \chi < \chi_{\rm cr}$ to the
asymptoticaly approached $\propto \chi^2$ continuum at $\chi > \chi_{\rm KN}$.
Such spectra are shown in Figure (\ref{KN-1}-\ref{KN-2}) for $q=1$ and $q=2$, respectively,
assuming monoenergetic injection with $\chi_{\rm inj}=1$, 
$\chi_1 = 10^{-2}$, $\chi_{\rm KN} = 10^4$, and $\chi_2 = 10^{8}$. In each 
figure we use two different acceleration timescales (thick solid and dashed
lines), but the radiative losses timescale, $t_{\rm loss}$, as
well as the normalization of the injection function, $\int dp \,
\widetilde{Q}(p)$, are kept constant. The emerging spectra are compared with the
electron energy distribution $\widetilde{N}(\chi)$ corresponding to the same
injection and cooling conditions, but with the momentum diffusion neglected
(equations \ref{kardashev}; thin solid lines in the lower panels of the
figures).

\begin{figure}
\centering
\includegraphics[scale=1.6]{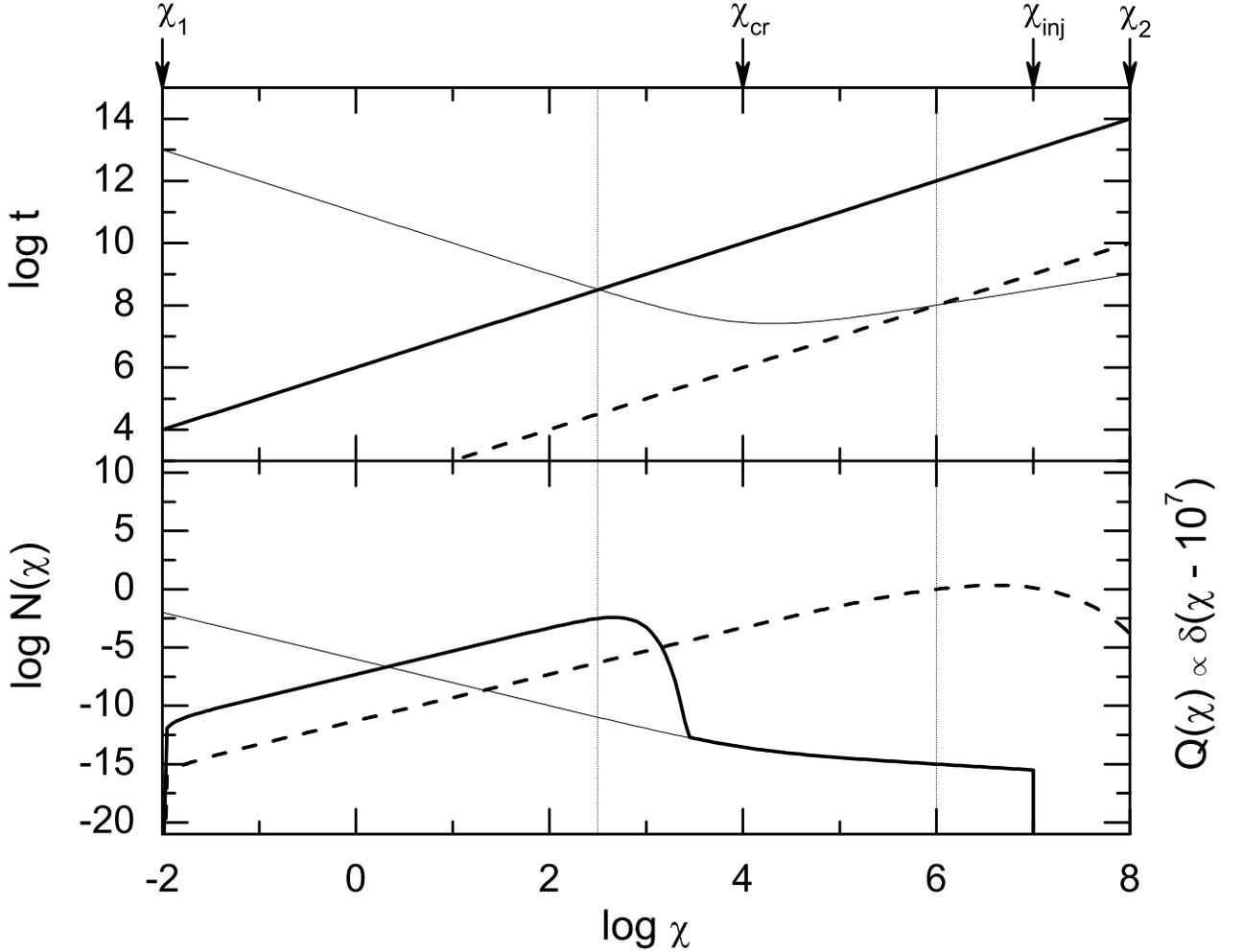}
\caption{{\it Upper panel:} Stochastic acceleration timescales for fixed $q=1$
and different plasma parameters (thick solid and dashed lines). Thin solid line
denotes inverse-Compton energy losses timescale considered with the assumed
$\chi_{\rm cr} = 10^4$. {\it Lower panel:} Particle spectra resulting from joint
stochastic acceleration and inverse-Compton energy losses specified in the upper
panel. The spectra correspond to the monoenergetic injection $Q(\chi) \propto
\delta(\chi-10^7)$ with fixed $\int dp \, \widetilde{Q}(p)$, and no particle
escape. Thin solid line denotes particle spectrum expected for the same
injection and cooling conditions, but with the momentum diffusion effects
neglected, $\widetilde{N}(\chi)$.}
\label{KN-3}
\end{figure}

\begin{figure}
\centering
\includegraphics[scale=1.6]{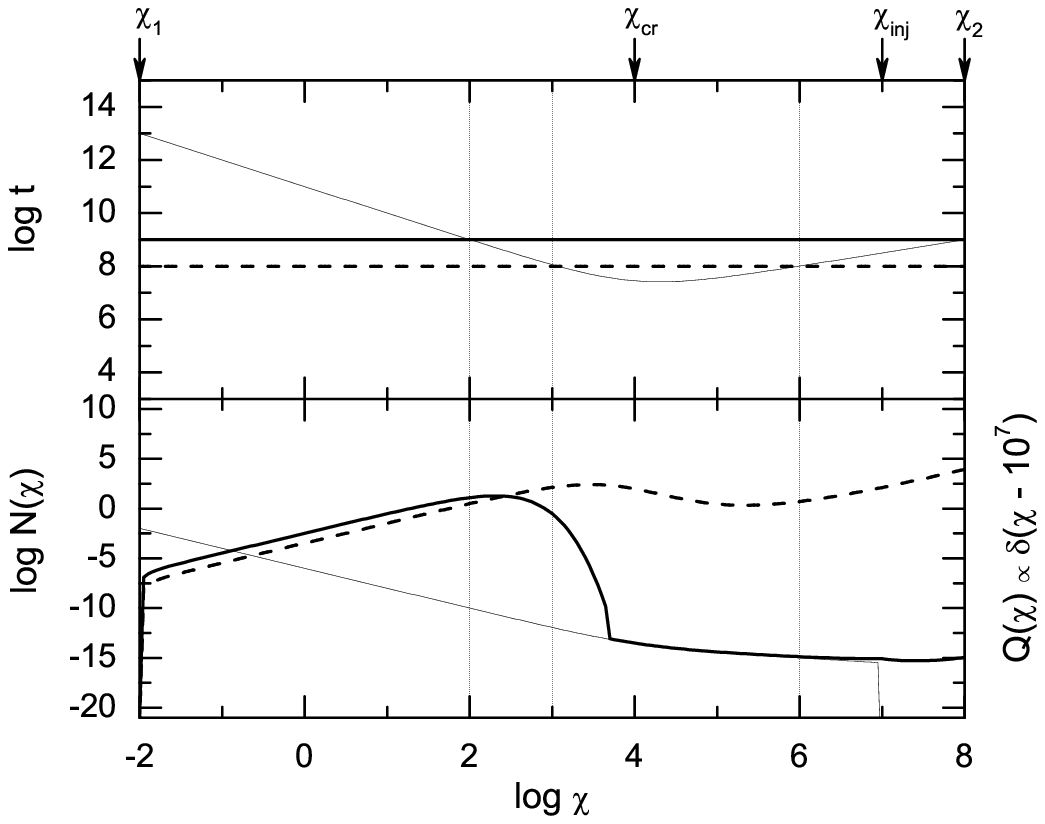}
\caption{The same as Figure (\ref{KN-3}) except for $q=2$.}
\label{KN-4}
\end{figure}

In the case of $q\neq 1.5$ and high-energy injection $\chi_{\rm inj} > \chi_{\rm cr}$, the
appropriate Green's fuction retains again familiar shape $\left.
\mathcal{G}(\chi, \chi_{\rm inj})\right|_{\rm ic; \, \chi_{\rm inj}>}^{q \neq 1.5} \sim {1 \over 1+q}
\, \chi_1^{-1-q} \, \chi^2 \, e^{- \, {1 \over 3-q} \, \left(\chi / \chi_{\rm
T}\right)^{3-q}}$ at low particle momenta $\chi < \chi_{\rm cr}$. And again, at
$\chi > \chi_{\rm cr}$ significant deviations from such a form may be observed,
as follows from the approximate form of the Green's function 
\begin{eqnarray}
& & \left. \mathcal{G}(\chi, \chi_{\rm inj})\right|_{\rm ic; \, \chi_{\rm inj}>}^{q \neq 1.5} \approx
\chi^2 \, e^{ - {1 \over 1.5-q} \left(\chi / \chi_{\rm KN}\right)^{1.5-q}} \,
e^{- \, {2 \over \sqrt{\pi}} \, \Gamma(3-q) \, \Gamma(q-1.5) \, \left(\chi_{\rm
cr} / \chi_{\rm Th}\right)^{3-q}} \times \label{green-ic-high} \\
& & \times \int_{\chi_1}^{\min(\chi_{\rm inj}, \chi)} d\chi' \, \chi'^{-2-q} \, \exp\left\{ {1
\over 3-q} \left({\chi' \over \chi_{\rm Th}}\right)^{3-q} \, F\!\left[{3 \over
2}, \, 3 - q,\, 4-q,\, -{\chi' \over \chi_{\rm cr}}\right]\right\} \quad {\rm
for} \quad \chi > \chi_{\rm cr} \nonumber
\end{eqnarray}
\noindent
(see equation \ref{green-ic} with the $A^{-1}$ term neglected). The resulting
particle spectra are plotted in Figures (\ref{KN-3}-\ref{KN-4}), where we
consider two limiting cases of $q=1$ and $q=2$, and assume monoenergetic
injection $Q(\chi) \propto \delta(\chi-\chi_{\rm inj})$ with $\chi_{\rm inj} =
10^7$. All the other parameters are fixed as before. As shown, in addition to
the spectral features discussed in the previous paragraph for the case of a
low-energy injection (Figures \ref{KN-1}-\ref{KN-2}), the radiatively-cooled
continuum may be observed at high particles energies $\chi < \chi_{\rm inj}$, depending on
the efficiency of the acceleration process. The KN effects manifest thereby by
means of a characteristic spectral flattening over the `standard' power-law form
$\propto \chi^{-2}$, obviously only within the momentum range $\chi_{\rm cr} <
\chi < \chi_{\rm inj}$, in agreement with the appropriate $\widetilde{N}(\chi)$
distribution (thin solid lines in the lower panels of Figures
\ref{KN-3}-\ref{KN-4}). Such a feature, being a direct result of a dominant
IC/KN-regime radiative cooling with the momentum diffusion effects negligible,
was discussed previously by, e.g., \citet{kus05,mod05,man07}.

\begin{figure}
\centering
\includegraphics[scale=1.6]{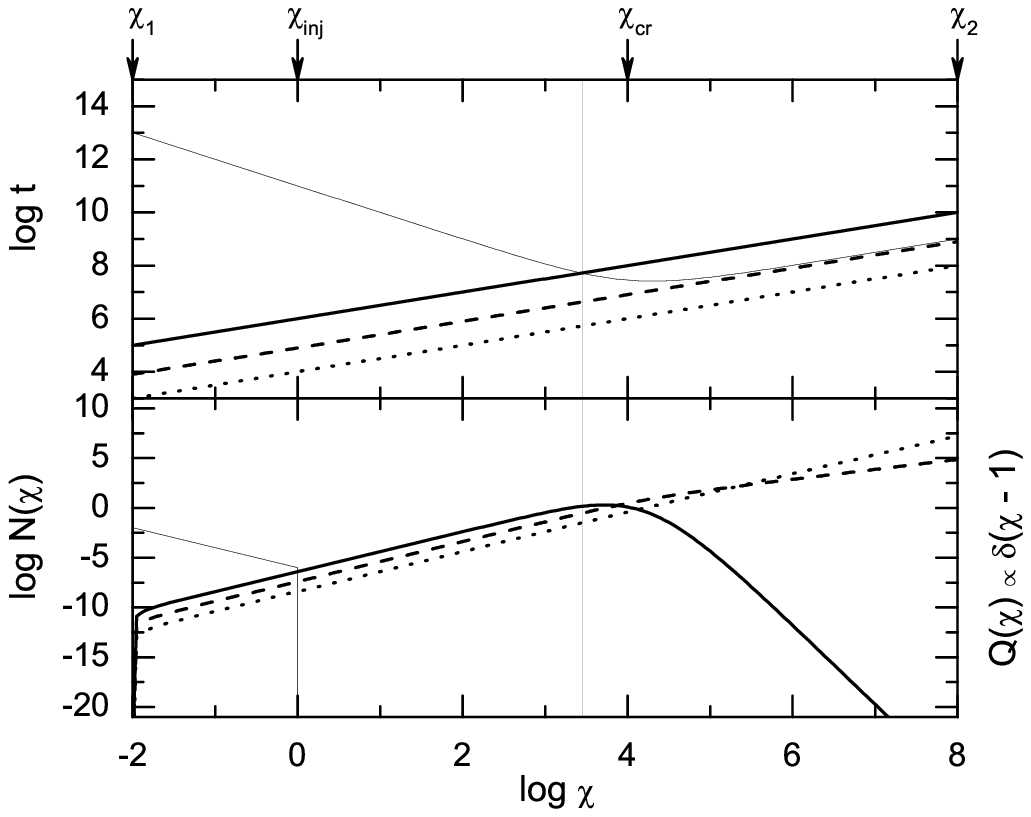}
\caption{{\it Upper panel:} Stochastic acceleration timescales for fixed $q=3/2$
and different plasma parameters (thick solid, dashed, and dotted lines). Thin
solid line denotes inverse-Compton energy losses timescale considered with the
assumed $\chi_{\rm cr} = 10^4$. {\it Lower panel:} Particle spectra resulting
from joint stochastic acceleration and inverse-Compton energy losses specified
in the upper panel. The spectra correspond to the monoenergetic injection
$Q(\chi) \propto \delta(\chi-1)$ with fixed $\int dp \, \widetilde{Q}(p)$, and
no particle escape. Thin solid line denotes particle spectrum expected for the
same injection and cooling conditions, but with the momentum diffusion effects
neglected, $\widetilde{N}(\chi)$.}
\label{KN-5}
\end{figure}

Finally, for completeness we note that with $q = 1.5$ one can solve
equation (\ref{s-loss}) to obtain
\begin{equation}
S(\chi) = \chi^{-2} \exp\left\{ 2 \, \left({\chi_{\rm cr} \over \chi_{\rm
T}}\right)^{3/2} \, \left({\rm ArcSinh}\!\sqrt{\chi / \chi_{\rm cr}} -
{\sqrt{\chi / \chi_{\rm cr}} \over \sqrt{1 + (\chi / \chi_{\rm cr})}}
\right)\right\} \, .
\label{krei}
\end{equation}
This reduces to $S(\chi) \sim \chi^{-2} \, \exp\left[{2\over 3} \left(\chi /
\chi_{\rm Th}\right)^{3/2}\right]$ for $\chi < \chi_{\rm cr}$, and can be
approximated by $S(\chi) \sim 0.54 \, \chi^{-1} \, \chi_{\rm cr}^{-1}$ for $\chi
> \chi_{\rm cr}$. The resulting particle spectra, shown in Figure (\ref{KN-5})
for the case of a low-energy injection $Q(\chi) \propto \delta(\chi - 1)$, are
therefore $N(\chi<\chi_{\rm cr}) \propto \chi^{2} \, \exp\left[-{2\over 3}
\left(\chi / \chi_{\rm Th}\right)^{3/2}\right]$ at low momenta, or of the
power-law form $N(\chi>\chi_{\rm cr}) \propto \chi^{-\sigma'}$ at higher momenta
where the KN effects are important. Here $\sigma' \equiv {t_{\rm acc} \over
t_{\rm ic}(\chi > \chi_{\rm cr})} - 2 = {\tau_{\rm acc} \over \tau_{\rm ic}} \,
\chi_{\rm cr}^{1.5} -2 $. 

\subsection{Bremsstrahlung and Coulomb Energy Losses}

At high densities or low magnetic field (in general low Alfv{\'e}n velocities) 
electron-electron and electron-ion interactions become important. These result in an 
elastic loss due to Coulomb collisions or radiative loss via bremsstrahlung. 
At low energies the bremsstrahlung loss rate is negligible when compared  to 
the Coulomb loss rate, which is independent of energy for 
relativistic charge particles \citep[see e.g.][]{pet73,pet01}. However, since the 
bremsstrahlung rate increases nearly linearly with energy, above some critical 
energy bremsstarahlung becomes dominant. The time scales associated with these 
processes approximately are 
\begin{equation}
t_{\rm coul} = \tau_{\rm coul} \, \chi , \quad {\rm where} \quad \tau_{\rm coul}
\equiv {p_0 \over m_{\rm e} c} \, {2 \over 3 \, \sigma_{\rm Th} c \, n_{\rm g} \, \ln\Lambda}
\label{coul}
\end{equation}
\noindent
and
\begin{equation}
t_{\rm brem} = \tau_{\rm brem} \, , \quad {\rm where} \quad \tau_{\rm brem}
\equiv {\pi \over 3 \, \alpha_{\rm fs} \sigma_{\rm Th} c \, n_{\rm g}} \, .
\label{brem}
\end{equation}
\noindent
Here $n_{\rm g}$ is the background plasma density, the Coulomb logarithm $\ln\Lambda$ 
varies from 10 to 40 for variety of astrophysical plasma, $\alpha_{\rm fs}=1/137$ is 
the fine structure constant, and the bremsstrahlung rate includes electron-ion and 
electron-electron bramsstrahlung, and assumes completely unscreened limit with 
approximately $10\%$ (fully ionized) helium abundance \citep{blu70}.
The time scales are equal at energy $p_{\rm Coul} = \pi \, \ln \Lambda \, m_{\rm e} c/
(2 \, \alpha_{\rm fs})$. At higher energies the bremsstrahlung loss becomes unimportant 
compared to the synchrotron or IC losses. For example, the synchrotron loss becomes 
equal to and exceeds the bremsstrahlung loss at electron momenta $p\geq p_{\rm brem}
\equiv (m_e/m_p)(\alpha_{\rm fs}/\beta_{\rm A}^2) \, m_{\rm e} c$ so that for bremsstrahlung to be at 
all important we need $1000<p/(m_ec)< 10^{-5} \, \beta_{\rm A}^{-2}$, requiring $\beta_{\rm A}<0.003$. 
Below we investigate in some details stochastic acceleration for the conditions 
when the Coulomb and bremsstrahlung processes are the dominant loss processes.
 
At low energies, $p<p_{\rm Coul}$, Coulomb collision dominate. If $p_0\gg m_ec$ then 
in the range $m_ec\ll p\ll p_{\rm Coul}$ and for $q > 1$, the appropriate Green's function becomes
\begin{eqnarray}
& & \left. \mathcal{G}(\chi, \chi_{\rm inj})\right|_{\rm coul}^{q>1} = \chi^2 \, e^{{1 \over
1-q} \, \left[\left({\chi_1 \over \chi_{\rm eq}}\right)^{1-q} - \left({\chi
\over \chi_{\rm eq}}\right)^{1-q}\right]} \, \left({1 \over A} +
\int^{\min[\chi_{\rm inj},\, \chi]}_{\chi_1} d\chi' \, \chi'^{-(2+q)} \, e^{{1 \over 1-q}
\, \left[ \left({\chi' \over \chi_{\rm eq}}\right)^{1-q}- \left({\chi_1 \over
\chi_{\rm eq}}\right)^{1-q}\right]} \right) \approx \nonumber \\
& & \approx \chi^2 \, e^{- \, {1 \over 1-q} \, \left({\chi \over \chi_{\rm
eq}}\right)^{1-q}} \, {\chi_{\rm eq}^{-1-q} \, (-1)^{2 / (1-q)} \over (1-q)^{2 /
(1-q)}} \, \Gamma\left[ - {1+q \over 1-q} \, , \, - \, {\left(\min[\chi_{\rm inj},\, \chi]
/ \chi_{\rm eq}\right)^{1-q} \over 1-q} , \, - \, {\left(\chi_1 / \chi_{\rm
eq}\right)^{1-q} \over 1-q} \right] \label{green-coul}
\end{eqnarray}
\noindent
(see equations \ref{green-loss}$-$\ref{s-loss}), where the equilibrium momentum
$\chi_{\rm eq}= (\tau_{\rm coul}/\tau_{\rm acc})^{1/(1-q)}$ is defined by the
$t_{\rm acc} = t_{\rm coul}$ condition, yielding $\vartheta_{\chi} = \chi_{\rm
eq}^{q-1}/\chi$. Note that since $q>1$ are considered, the acceleration
timescale is longer than the Coulomb interactions timescale for $\chi <
\chi_{\rm eq}$. Thus, in the case of a low-energy particle injection with
$\chi_{\rm inj} < \chi_{\rm eq}$, the emerging particle spectra are of the
`cooled' form $N(\chi) = \widetilde{N}(\chi) \propto const$ (see equation
\ref{kardashev} with $\vartheta_{\chi} \propto \chi^{-1}$). If, however,
higher-energy particles are injected to the system, an additional flat-spectrum
component $N(\chi) \propto \chi^2$ is formed at $\chi > \chi_{\rm eq}$.

Let us finally note, that pure Coulomb energy losses and the Bohm limit $q=1$
correspond to the situation when $\vartheta_{\chi} = const$, and hence $S(\chi)
= \chi^{-2 + (\tau_{\rm acc}/\tau_{\rm coul})}$. The Green's function
(\ref{green-loss}) adopts then the form
\begin{eqnarray}
\left. \mathcal{G}(\chi, \chi_{\rm inj})\right|_{\rm coul}^{q=1} = \chi^{2-{\tau_{\rm acc}
\over \tau_{\rm coul}}} \, \left({1 \over A} + \int^{\min[\chi_{\rm inj},\, \chi]}_{\chi_1}
d\chi' \, \chi'^{-4+{\tau_{\rm acc} \over \tau_{\rm coul}}} \right) \sim
\nonumber \\
\sim {1 \over \sigma'} \, \chi^{-\sigma'} \, \left\{ \begin{array}{ccc}
\chi_1^{\sigma'} \quad & {\rm for} & \tau_{\rm acc}/\tau_{\rm coul} < 2 \\
\min^{\sigma'}(\chi_{\rm inj}, \,\chi) \quad & {\rm for} & \tau_{\rm acc}/\tau_{\rm coul} >
2 \end{array} \right. \, , \label{green-coul2}
\end{eqnarray}
\noindent
where $\sigma' \equiv {\tau_{\rm acc} \over \tau_{\rm coul}} - 2$. Hence, if
only $\tau_{\rm acc} < 2 \, \tau_{\rm coul}$, a power-law particle energy
distribution $N(\chi) \propto \chi^{-\sigma'}$ forms, with $-2 < \sigma' < 0$.
For any longer acceleration timescale, $\tau_{\rm acc} > 2 \, \tau_{\rm coul}$,
and for the source function $Q(\chi) \propto \delta (\chi - \chi_{\rm inj})$,
the emerging electron spectra are $N(\chi) \propto const$ for $\chi < \chi_{\rm
inj}$, and $N(\chi) \propto \chi^{-\sigma'}$ with $\sigma' > 0$ for $\chi >
\chi_{\rm inj}$. This is consistent with the solution found by \citet{bog85},
who considered also synchrotron emission and finite escape timescale in addition
to the Coulomb energy losses of ultrarelativistic electrons interacting with
flat-spectrum turbulence $q=1$.

At higher energies and in the range $p_{\rm Coul}\ll p \ll p_{\rm brem}$ bremsstrahlung 
loss is the dominant process and  the equilibrium momentum defined by the condition
$t_{\rm acc} = t_{\rm brem}$ for $q<2$ becomes $\chi_{\rm eq}= (\tau_{\rm
brem}/\tau_{\rm acc})^{1/(2-q)}$, yielding $\vartheta_{\chi} = \chi_{\rm
eq}^{-(2-q)}$. Hence, the Green's function (\ref{green-loss}) is
\begin{eqnarray}
& & \left. \mathcal{G}(\chi, \chi_{\rm inj})\right|_{\rm brem}^{q<2} = \chi^2 \, e^{- \, {1
\over 2-q} \, \left({\chi \over \chi_{\rm eq}}\right)^{2-q}} \, \left({1 \over
A} + \int^{\min[\chi_{\rm inj},\, \chi]}_{\chi_1} d\chi' \, \chi'^{-(2+q)} \, e^{{1 \over
2-q} \, \left({\chi' \over \chi_{\rm eq}}\right)^{2-q}} \right) \approx \label{green-brem} \\
& & \approx \chi^2 \, e^{- \, {1 \over 2-q} \, \left({\chi \over \chi_{\rm
eq}}\right)^{2-q}} \, {\chi_{\rm eq}^{-1-q} \, (-1)^{3 / (2-q)} \over (2-q)^{3 /
(2-q)}} \, \Gamma\left[ - {1+q \over 2-q} \, , \, - \, {\left(\min[\chi_{\rm inj},\, \chi]
/ \chi_{\rm eq}\right)^{2-q} \over 2-q} , \, - \, {\left(\chi_1 / \chi_{\rm
eq}\right)^{2-q} \over 2-q} \right] \nonumber
\end{eqnarray}
\noindent
(equations \ref{green-loss}$-$\ref{s-loss}). In other words, for any injection
conditions the expected electron energy distribution is of the $N(\chi) \propto
\chi^2 \, \exp\left[- {1 \over 2-q} \left(\chi / \chi_{\rm
eq}\right)^{2-q}\right]$ form, except for the case when high energy particles
with $\chi_{\rm inj} > \chi_{\rm eq}$ are injected to the system. Such high
energy particles subjected to the bremsstrahlung energy losses form then an
additional `cooled' high-energy power-law tail $N(\chi) \propto \chi^{-1}$ in
the momentum range between $\chi_{\rm eq}$ and $\chi_{\rm inj}$, in agreement
with the appropriate form of $\widetilde{N}(\chi)$ with $\vartheta_{\chi} =
const$ (see equation \ref{kardashev}).

The situation changes for $q=2$, since both the acceleration and cooling
timescales are now independent of electrons' energy. In this case $S(\chi) =
\chi^{-2 + (\tau_{\rm acc}/\tau_{\rm brem})}$, and the Green's function
(\ref{green-loss}) adopts the form
\begin{eqnarray}
\left. \mathcal{G}(\chi, \chi_{\rm inj})\right|_{\rm brem}^{q=2} = \chi^{2-{\tau_{\rm acc}
\over \tau_{\rm brem}}} \, \left({1 \over A} + \int^{\min[\chi_{\rm inj},\, \chi]}_{\chi_1}
d\chi' \, \chi'^{-4+{\tau_{\rm acc} \over \tau_{\rm brem}}} \right) \sim \nonumber \\
\sim {1 \over 1-\sigma'} \, \chi^{-\sigma'} \, \left\{ \begin{array}{ccc}
\chi_1^{-1+\sigma'} \quad & {\rm for} & \tau_{\rm acc}/\tau_{\rm brem} < 3 \\
\min^{-1+\sigma'}(\chi_{\rm inj}, \,\chi) \quad & {\rm for} & \tau_{\rm acc}/\tau_{\rm
brem} > 3 \end{array} \right. \, , \label{green-brem3}
\end{eqnarray}
\noindent
where $\sigma' \equiv {\tau_{\rm acc} \over \tau_{\rm brem}}-2$. This is
consistent with the appropriate Green's function found by \citet{sch87} who, in
a framework of the `hard-sphere' approximation $q=2$, considered also
synchrotron emission and particle escape in addition to the bremsstrahlung
radiation. The solution (\ref{green-brem3}) implies that within the whole energy
range the expected electron energy distribution is of the power-law form
$N(\chi) \propto \chi^{-\sigma'}$, with the power-law index $-2 < \sigma' < 1$.
For any longer acceleration timescale, $\tau_{\rm acc} > 3 \, \tau_{\rm brem}$,
and monoenergetic injection $Q(\chi) \propto \delta (\chi - \chi_{\rm inj})$,
the emerging electron spectra are expected to be of the $N(\chi) \propto
\chi^{-1}$ form for $\chi < \chi_{\rm inj}$, while $N(\chi) \propto
\chi^{-\sigma'}$ with $\sigma' > 1$ for $\chi > \chi_{\rm inj}$.

\section{Efficient Particle Escape}

In this section we investigate steady-state solutions to the momentum diffusion
equation of radiating ultrarelativistic particles with a finite escape timescale
(equation \ref{steady}). Our analytical approach force us to consider only the
limiting cases of turbulent spectral indices $q=2$ or $q=1$, as well as to
restrict the analysis of radiative losses to the synchrotron and/or IC-Thompson 
regime processes, ($\vartheta_{\chi} \propto \chi$). We note that the global
approximation to the solution of the momentum diffusion equation not necessarily 
restricted to some particular values of the $q$ parameter, with the regular energy 
losses and particle escape terms included, were studied by \citet{gal95} by using
the WKBJ method. Just us before, we consider finite energy
range of particles undergoing momentum diffusion, $0 < \chi_1, \, \chi_2 <
\infty$, strictly related to the finite wavelength range of interacting
turbulent modes. We construct the Green's function
accordingly to the procedure outlined in the previous section \S\,3, with 
addition of the escape term ($\varepsilon\neq 0$) and 
with a different boundary conditions. Specifically, we change 
equation (\ref{eqfinal}) to
\begin{equation}
{\partial \mathcal{N} \over \partial \tau} + \left. \mathcal{F}\right|_{\chi_2}
- \left. \mathcal{F}\right|_{\chi_1} = \int_{\chi_1}^{\chi_2} d\chi \, Q(\chi,
\tau) - \varepsilon \, \int_{\chi_1}^{\chi_2} d\chi \, \chi^{2-q} \, N(\chi) \,
,
\label{cont-esc}
\end{equation}
\noindent
where the particle flux in the momentum space $\mathcal{F}[N(\chi)]$ is defined
in the same way as previously (equation \ref{flux1}). As evident,  the no-flux
boundary conditions, $\mathcal{F}[N(\chi_1)] = \mathcal{F}[N(\chi_2)] = 0$, and
conservation of total number of particles, $\partial \mathcal{N} / \partial \tau
=0$, (within the energy range $[\chi_1, \chi_2]$) implies that the particle injection is 
completely balanced by the particle escape. We will assume this to be the case 
in this section. Physical realization of these would imply
presence of an another efficient yet unspecified acceleration process operating
at $\chi < \chi_1$, which prevent negative particle momentum flux through the
$\chi_1$ boundary. As shown below, the solutions we obtain  agree with the ones
discussed in the literature for singular boundary conditions for
the infinite momentum range \citep{jon70,sch84,bog85,par95}, as long as we are dealing with particle
momenta $\chi \gg \chi_1$ and $\chi \ll \chi_2$.

\subsection{`Hard-Sphere' Approximation}

`Hard-Sphere' approximation for the momentum diffusion of ultrarelativistic
electrons undergoing synchrotron energy losses corresponds to the fixed $q=2$
and $\vartheta_{\chi} = \chi/\chi_{\rm eq}$ (see equations \ref{timescales} and
\ref{syn}). With these, the equation (\ref{steady}) adopts the form
\begin{equation}
\chi^2 \, N''(\chi) + \chi_{\rm eq}^{-1} \, \chi^2 \, N'(\chi) + \left(2 \,
\chi_{\rm eq}^{-1} \, \chi - 2 - \varepsilon\right) \, N(\chi) = -Q(\chi) \, .
\label{hss}
\end{equation}
\noindent
The two linearly-independent particular solutions to the homogeneous form of the
above equation are
\begin{eqnarray}
y_1(\chi) & = & \chi^{\sigma+1} \, e^{-{\chi \over \chi_{\rm eq}}} \,
U\!\left[\sigma-1, \, 2 \sigma +2, \, {\chi \over \chi_{\rm eq}}\right] \, ,
\nonumber \\
y_2(\chi) & = & \chi^{\sigma+1} \, e^{-{\chi \over \chi_{\rm eq}}} \,
M\!\left[\sigma-1, \, 2 \sigma +2, \, {\chi \over \chi_{\rm eq}}\right] \, ,
\label{hss-y}
\end{eqnarray}
\noindent
where $U[a,b,z]$ and $M[a,b,z]$ are Tricomi and Kummer confluent
hypergeometrical functions, respectively, and $\sigma \equiv -(1/2) + [(9/4) +
\varepsilon]^{1/2}$. Introducing next their linear combinations, $u_1(\chi) =
y_1(\chi) + \alpha \, y_2(\chi)$ and $u_2(\chi) = y_1(\chi) + \beta \,
y_2(\chi)$, one may find that the no-flux boundary conditions
$\mathcal{F}[u_1(\chi_1)]=\mathcal{F}[u_2(\chi_2)]=0$ are fulfilled for
\begin{equation}
\alpha = (2+\sigma) \, {U\!\left[\sigma, \, 2 \sigma +2 , \, \chi_1 / \chi_{\rm
eq} \right] \over M\!\left[\sigma, \, 2 \sigma +2 , \, \chi_1 / \chi_{\rm eq}
\right]} \, , \quad {\rm and} \quad
\beta = (2+\sigma) \, {U\!\left[\sigma, \, 2 \sigma +2 , \, \chi_2 / \chi_{\rm
eq} \right] \over M\!\left[\sigma, \, 2 \sigma +2 , \, \chi_2 / \chi_{\rm eq}
\right]} \, .
\label{hss-bc} 
\end{equation}
\noindent
This gives the Green's function of the problem as
\begin{eqnarray}
& & \left. \mathcal{G}(\chi, \chi_{\rm inj})\right|_{\rm esc}^{q=2} = {\Gamma(\sigma-1) \over
\Gamma(2\sigma+2)} \, \left(\alpha - \beta\right)^{-1} \, \chi_{\rm inj}^{-2} \, \chi_{\rm
eq}^{-2 \sigma - 1} \, e^{\chi_{\rm inj} /  \chi_{\rm eq}} \times \nonumber \\
& & \times \quad \left\{ \begin{array}{ccc}
\left[y_1(\chi) + \alpha \, y_2(\chi)\right] \, \left[y_1(\chi_{\rm inj}) + \beta \,
y_2(\chi_{\rm inj})\right] & {\rm for} & \chi_1 \leq \chi < \chi_{\rm inj} \\
\left[y_1(\chi_{\rm inj}) + \alpha \, y_2(\chi_{\rm inj})\right] \, \left[y_1(\chi) + \beta \,
y_2(\chi)\right] & {\rm for} & \chi_{\rm inj} < \chi \leq \chi_2
\end{array} \right. \, . \label{hss-G} 
\end{eqnarray}
\noindent

\begin{figure}
\centering
\includegraphics[scale=1.6]{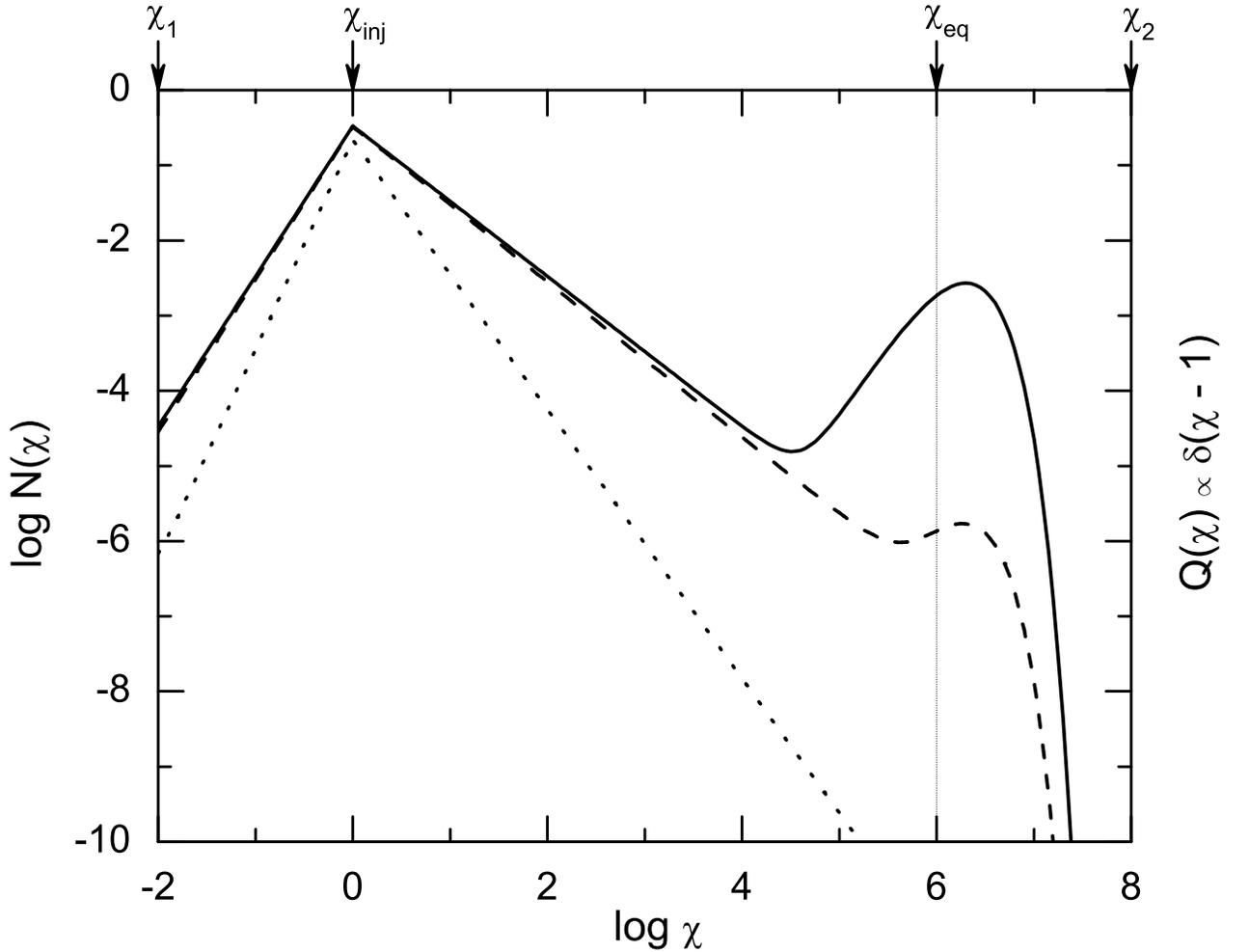}
\caption{`Hard-sphere approximation' ($q=2$): particle spectra resulting from
joint stochastic acceleration, particle escape, and synchrotron energy losses.
The spectra correspond to the monoenergetic injection $Q(\chi) \propto
\delta(\chi-\chi_{\rm inj})$ with fixed normalization, fixed acceleration and
cooling rates, but different escape timescales (parameter $\varepsilon = 3$,
$0.1$, $10^{-4}$; dotted, dashed, and solid lines, respectively). For
illustration, $\chi_1 = 10^{-2}$, $\chi_{\rm inj} = 1$, $\chi_{\rm eq} = 10^6$,
and $\chi_2 = 10^{8}$ have been selected.
}
\label{hss1}
\end{figure}

\begin{figure}
\centering
\includegraphics[scale=1.6]{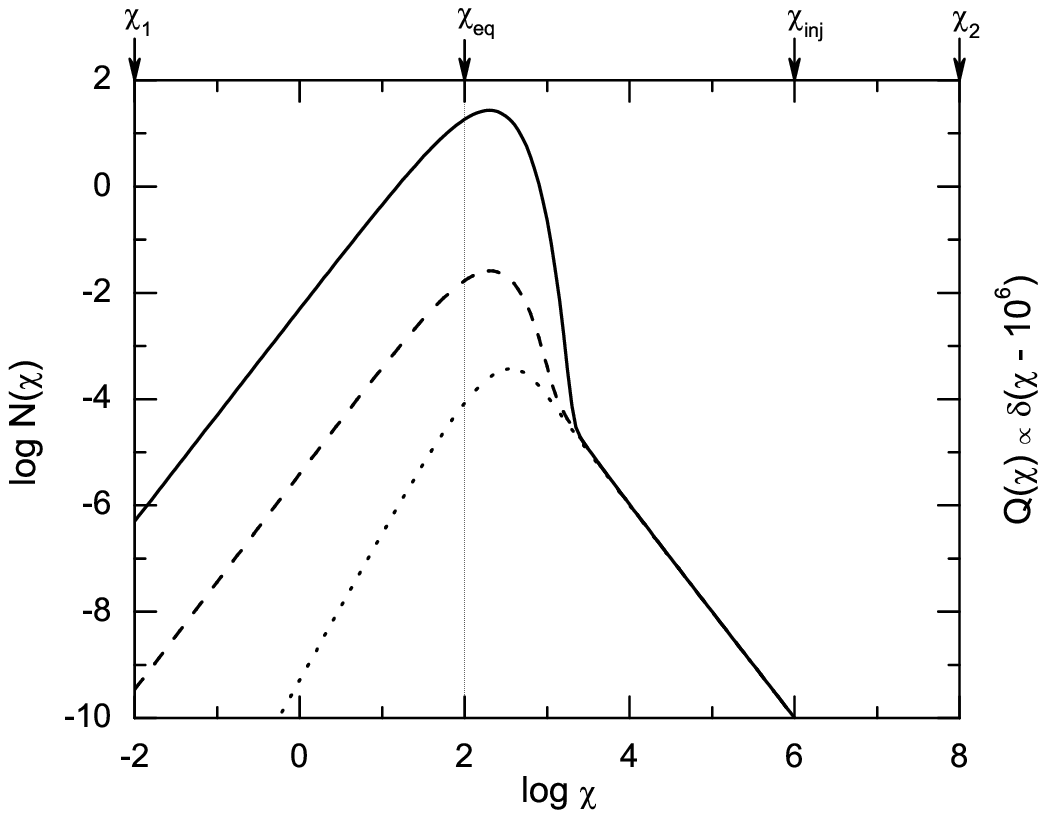}
\caption{The same as FIgure (\ref{hss1}) except for $\chi_{\rm inj} = 10^6$.}
\label{hss2}
\end{figure}

In order to investigate the above solution, let us consider first the case
$\chi_1 \ll \chi_{\rm inj} \ll \chi_{\rm eq} \ll \chi_2$, and use the standard expansion of
the confluent hypergeometrical functions: $U[a,b,z] \sim z^{-a}$ and $M[a,b,z]
\sim \Gamma(b) \, e^z \, z^{a-b} / \Gamma(a)$ for $z \rightarrow \infty$, while
$U[a,b,z] \sim \Gamma(b-1) \, z^{1-b} / \Gamma(a)$ and $M[a,b,z] \sim 1$ for $z
\rightarrow 0$ \citep{abr64}. In this limit one gets
\begin{equation}
\left. \mathcal{G}(\chi, \chi_{\rm inj})\right|_{\rm esc, \, \chi_{\rm inj}<}^{q=2} \sim 
\left\{ \begin{array}{ccc}
{1 \over 2 \sigma + 1} \, \chi_{\rm inj}^{-\sigma-2} \, \chi^{\sigma+1}
& {\rm for} & \chi_1 < \chi < \chi_{\rm inj} \\
{1 \over 2 \sigma + 1} \, \chi_{\rm inj}^{\sigma-1} \, \chi^{-\sigma} 
& {\rm for} & \chi_{\rm inj} < \chi \ll \chi_{\rm eq} \\
{\Gamma(\sigma-1) \over \Gamma(2\sigma+2)} \, \chi_{\rm inj}^{\sigma-1}  \, \chi_{\rm
eq}^{-\sigma-2} \, \chi^2 \, e^{-\chi / \chi_{\rm eq}} & {\rm for} & \chi_{\rm
eq} \lesssim \chi < \chi_2
\end{array} \right. \, .
\label{hss-G-approx}
\end{equation}
\noindent
Thus, by moving the critical momenta $\chi_1$ and $\chi_2$ toward $0$ and
$\infty$, respectively, the resultant Green's function approaches
asymptotically --- as expected --- the corresponding Green's function 
for singular boundary conditions obtained by \citet{jon70,sch84} and \citet{par95}. 
In particular, one can find that with the monoenergetic injection $Q(\chi) \propto
\delta(\chi-\chi_{\rm inj})$, the resulting electron energy distribution is then
of the form $N(\chi<\chi_{\rm inj}) \propto \chi^{\sigma+1}$ and
$N(\chi>\chi_{\rm inj}) \propto \chi^{-\sigma}$ up to maximum momentum
$\chi_{\rm eq}$. Moreover, for the increasing escape timescale $\varepsilon
\rightarrow 0$, one has $\sigma \approx 1$ and the pile-up bump $N(\chi) \propto
\chi^2 \, \exp\left[-\chi / \chi_{\rm eq}\right]$ emerging around $\chi \sim
\chi_{\rm eq}$ energies. This is shown in Figure (\ref{hss1}), where we fixed
normalization of the monoenergetic injection $\int dp \, \widetilde{Q}(p)$,
acceleration and losses timescales, but varied the escape timescale
($\varepsilon = 3$, $0.1$, $10^{-4}$; dotted, dashed, and solid lines,
respectively). For illustration we have selected $\chi_1 = 10^{-2}$, $\chi_{\rm
inj} = 1$, $\chi_{\rm eq} = 10^6$, and $\chi_2 = 10^{8}$.

When high energy particles are injected to the system, such that $\chi_1 \ll
\chi_{\rm eq} \ll \chi_{\rm inj} \ll \chi_2$, one may find useful asymptotic expansion of
the Green's function
\begin{equation}
\left. \mathcal{G}(\chi, \chi_{\rm inj})\right|_{\rm esc, \, \chi_{\rm inj}>}^{q=2} \sim 
\left\{ \begin{array}{ccc}
{\Gamma(\sigma-1) \over \Gamma(2 \sigma + 2)} \, \chi_{\rm eq}^{-\sigma-2} \,
\chi^{\sigma+1} \, e^{-\chi/\chi_{\rm eq}} 
& {\rm for} & \chi_1 < \chi \lesssim \chi_{\rm eq} \\
\chi^{-2} \, \chi_{\rm eq} & {\rm for} & \chi_{\rm eq} \ll \chi < \chi_{\rm inj} \\
\chi_{\rm inj}^{-4} \, \chi_{\rm eq} \, e^{\chi_{\rm inj}/\chi_{\rm eq}} \, \chi^2 \,
e^{-\chi/\chi_{\rm eq}} & {\rm for} & \chi_{\rm inj} < \chi < \chi_2
\end{array} \right. \, .
\label{hss-G-approx2}
\end{equation}
\noindent
That is, for the monoenergetic injection $Q(\chi) \propto \delta(\chi-\chi_{\rm
inj})$ with $\chi_{\rm inj} > \chi_{\rm eq}$ the resulting electron energy
distribution is of the form $N(\chi) \propto \chi^{\sigma+1} \,
\exp\left[-\chi/\chi_{\rm eq}\right]$ for $\chi \lesssim \chi_{\rm eq}$.
However, within the energy range $\chi_{\rm eq} < \chi < \chi_{\rm inj}$ the
power-law tail $N(\chi) \propto \chi^{-2}$ emerges, representing radiatively
($\vartheta_{\chi} \propto \chi$) cooled high-energy particles injected to the
system, undergoing negligible (when compared to the energy loss rate) momentum
diffusion. At even higher energies, $\chi > \chi_{\rm inj}$, the particle
spectrum cuts-off rapidly. This is shown in Figure (\ref{hss2}), where, as
before, we fixed normalization of the monoenergetic injection $\int dp \,
\widetilde{Q}(p)$, acceleration and losses timescales, but varied the escape
timescale ($\varepsilon = 3$, $0.1$, $10^{-4}$; dotted, dashed, and solid lines,
respectively). For illustration we have selected $\chi_1 = 10^{-2}$, $\chi_{\rm
eq} = 10^2$, $\chi_{\rm inj} = 10^6$, and $\chi_2 = 10^{8}$. Note, that the
esape timescale, and hence parameter $\varepsilon$, influences now the slope and
normalization of particle energy distribution only in the `low-energy' regime
$\chi < \chi_{\rm eq}$, such that the spectrum approaches $\propto x^2$ for
$\varepsilon \rightarrow 0$.

\subsection{Bohm Limit}

Bohm limit for the momentum diffusion of ultrarelativistic electrons undergoing
synchrotron energy losses corresponds to  $q=1$ and $\vartheta_{\chi} =
\chi/\chi_{\rm eq}^2$ (see equations \ref{timescales} and \ref{syn}). The
difference with the `hard-sphere' approximation is that the balance between
acceleration and escape timescales, $t_{\rm acc} = t_{\rm esc}$, define now yet
another critical energy, $\chi_{\rm esc} = \varepsilon^{-1/2}$ and
equation (\ref{steady}) takes the form
\begin{equation}
\chi \, N''(\chi) + \left(\chi_{\rm eq}^{-2} \, \chi^2 -1\right) \, N'(\chi) +
\left(2 \, \chi_{\rm eq}^{-2} \, \chi - \chi_{\rm esc}^{-2} \, \chi\right) \,
N(\chi) = -Q(\chi) \, .
\label{bohm-syn}
\end{equation}
\noindent
The two linearly-independent particular solutions to the homogeneous form of the
above equation are
\begin{eqnarray}
y_1(\chi) & = & \chi^{2} \, e^{-{1 \over 2} \, \left({\chi \over \chi_{\rm
eq}}\right)^2} \, U\!\left[\eta, \, 2, \, {1 \over 2} \left({\chi \over
\chi_{\rm eq}}\right)^2\right] \, , \nonumber \\
y_2(\chi) & = & \chi^{2} \, e^{-{1 \over 2} \, \left({\chi \over \chi_{\rm
eq}}\right)^2} \, M\!\left[\eta, \, 2, \, {1 \over 2} \left({\chi \over
\chi_{\rm eq}}\right)^2\right] \, , \label{bohm-syn-y}
\end{eqnarray}
\noindent
where $\eta \equiv {1 \over 2} (\chi_{\rm eq} / \chi_{\rm esc})^2$. Defining
$u_1(\chi) = y_1(\chi) + \alpha \, y_2(\chi)$ and $u_2(\chi) = y_1(\chi) +
\beta \, y_2(\chi)$, one finds that the no-flux boundary conditions
$\mathcal{F}[u_1(\chi_1)]=\mathcal{F}[u_2(\chi_2)]=0$ corresponds to
\begin{equation}
\alpha = 2 \, {U\!\left[\eta+1, \, 3 , \, {1 \over 2} \left(\chi_1 / \chi_{\rm
eq}\right)^2 \right] \over M\!\left[\eta+1, \, 3 , \, {1 \over 2} \left(\chi_1 /
\chi_{\rm eq}\right)^2 \right]} \, , \quad {\rm and} \quad
\beta = 2 \, {U\!\left[\eta+1, \, 3 , \, {1 \over 2} \left(\chi_2 / \chi_{\rm
eq}\right)^2 \right] \over M\!\left[\eta+1, \, 3 , \, {1 \over 2} \left(\chi_2 /
\chi_{\rm eq}\right)^2 \right]} \, .
\label{bohm-syn-bc} 
\end{equation}
\noindent
This gives the Green's function of the problem as
\begin{eqnarray}
& & \left. \mathcal{G}(\chi, \chi_{\rm inj})\right|_{\rm esc}^{q=1} = 
{1 \over 4} \, \Gamma(\eta) \, \left(\alpha - \beta\right)^{-1} \, \chi_{\rm inj}^{-2} \,
\chi_{\rm eq}^{-2} \, e^{{1 \over 2} (\chi_{\rm inj} /  \chi_{\rm eq})^2} \times
\nonumber \\
& & \times \quad \left\{ \begin{array}{ccc}
\left[y_1(\chi) + \alpha \, y_2(\chi)\right] \, \left[y_1(\chi_{\rm inj}) + \beta \,
y_2(\chi_{\rm inj})\right] & {\rm for} & \chi_1 \leq \chi < \chi_{\rm inj} \\
\left[y_1(\chi_{\rm inj}) + \alpha \, y_2(\chi_{\rm inj})\right] \, \left[y_1(\chi) + \beta \,
y_2(\chi)\right] & {\rm for} & \chi_{\rm inj} < \chi \leq \chi_2
\end{array} \right. \, . \label{bohm-syn-G}
\end{eqnarray}
\noindent

\begin{figure}
\centering
\includegraphics[scale=1.6]{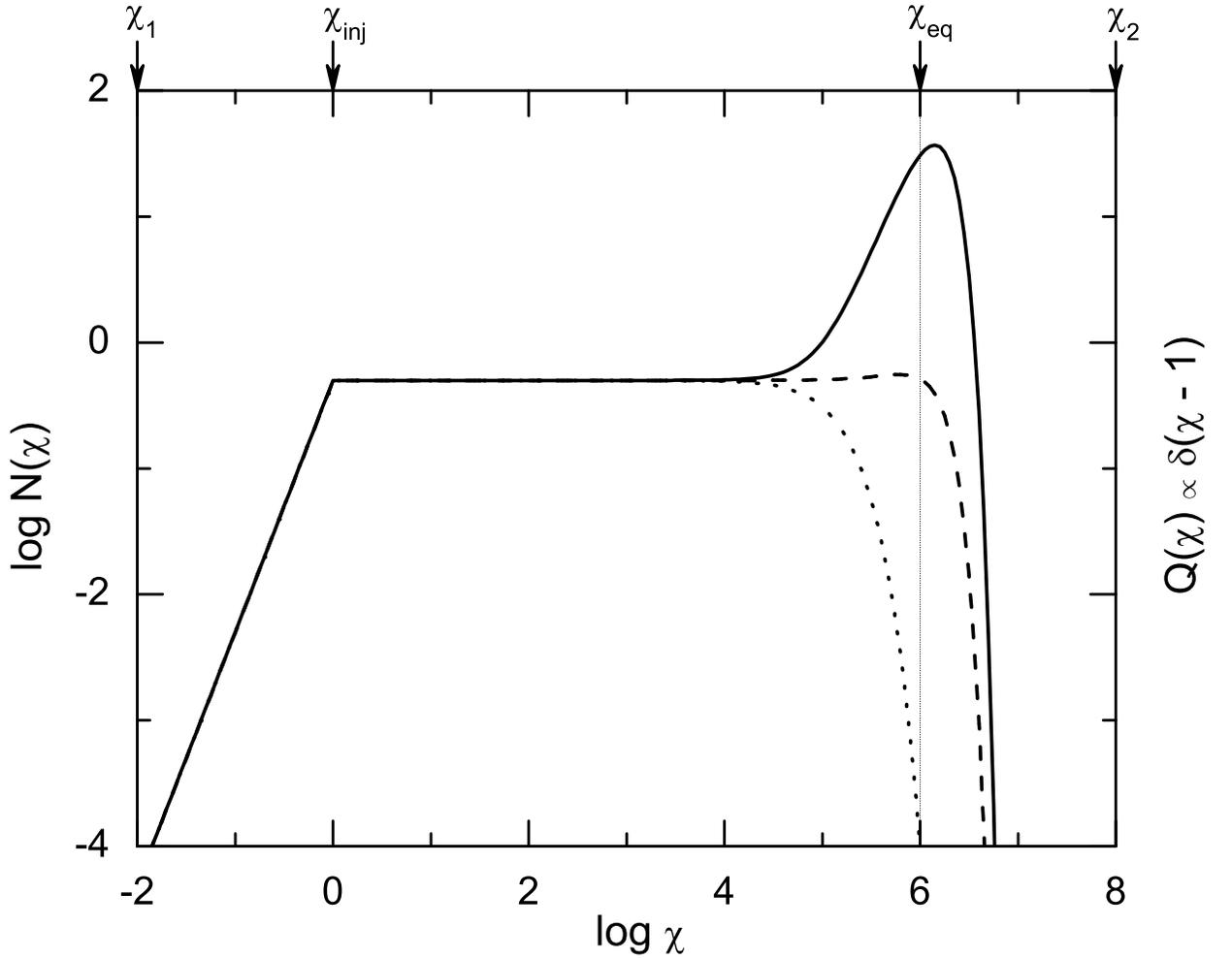}
\caption{Bohm Limit ($q=1$): particle spectra resulting from joint stochastic
acceleration, particle escape, and synchrotron energy losses. The spectra
correspond to the monoenergetic injection $Q(\chi) \propto \delta(\chi-\chi_{\rm
inj})$ with fixed normalization, fixed acceleration and cooling rates, but
different escape timescales (critical momenta $\chi_{\rm esc} = 10^{5}$, $10^6$,
$10^7$; dotted, dashed, and solid lines, respectively). For illustration,
$\chi_1 = 10^{-2}$, $\chi_{\rm inj} = 1$, $\chi_{\rm eq} = 10^6$, and $\chi_2 =
10^{8}$ have been selected.}
\label{bohm-syn-1}
\end{figure}

Let us consider first the case 
$\chi_1 \ll \chi_{\rm inj} \ll \chi_{\rm eq} \ll \chi_2$ for which he Green's function of equation 
(\ref{bohm-syn-G}) can be then approximated as
\begin{equation}
\left. \mathcal{G}(\chi, \chi_{\rm inj})\right|_{\rm esc, \, \chi_{\rm inj}<}^{q=1} \sim 
\left\{ \begin{array}{ccc}
{1 \over 2} \, \chi_{\rm inj}^{-2} \, \chi^{2} & {\rm for} & \chi_1 < \chi < \chi_{\rm inj} \\
{1 \over 2} & {\rm for} & \chi_{\rm inj} < \chi \ll \chi_{\rm eq} \\
2^{\eta -2} \, \Gamma(\eta) \, \chi_{\rm eq}^{-2+2 \, \eta} \, \chi^{2 - 2 \,
\eta} \, e^{-{1 \over 2} (\chi / \chi_{\rm eq})^2} & {\rm for} & \chi_{\rm eq}
\lesssim \chi < \chi_2 
\end{array} \right. \, .
\label{bohm-G-approx}
\end{equation}
\noindent
Note that, as expected, in the limits  $\chi_1 \rightarrow 0$ and $\chi_2 \rightarrow \infty$, 
the Green's function (\ref{bohm-syn-G}) approaches
asymptotically the solution obtained for singular boundary
conditions by \citet{bog85}. As shown in  Figure (\ref{bohm-syn-1}), for a monoenergetic injection $Q(\chi) \propto
\delta(\chi-\chi_{\rm inj})$,  the resulting electron energy distribution is
$N(\chi<\chi_{\rm inj}) \propto \chi^{2}$ and $N(\chi>\chi_{\rm inj}) \propto
const$ up to maximum momentum $\chi_{\rm eq}$, with the spectral indexes 
independent of the value of the escape timescale. However, for energies near 
and above $\chi_{\rm eq}$ the spectra depend on the value of $\eta$. For  
$\eta \rightarrow 0$, \i.e. when the escape timescale is large, the familiar bump $N(\chi) \propto
\chi^2 \, \exp\left[-{1 \over 2} (\chi / \chi_{\rm eq})^2\right]$ emerges around
$\chi \sim \chi_{\rm eq}$ energies (solid line). In the opposite case, when $\eta > 1$ (or
$\chi_{\rm eq} > \chi_{\rm esc}$), no pile-up bump is present, and the electron
spectrum cut-offs exponentially at $\chi_{\rm esc}$ momenta (dashed and dotted lines). 
Here, as before, we fixed the normalization of the
monoenergetic injection $\int \widetilde{Q}(p) \, dp$, and the acceleration and loss
timescales, but varied the escape timescale such that $\chi_{\rm esc} = 10^{5}$,
$10^6$, and $10^7$ (dotted, dashed, and solid lines, respectively). We choose
$\chi_1 = 10^{-2}$, $\chi_{\rm inj} = 1$, $\chi_{\rm eq} = 10^6$, and $\chi_2 =
10^{8}$. 

\begin{figure}
\centering
\includegraphics[scale=1.6]{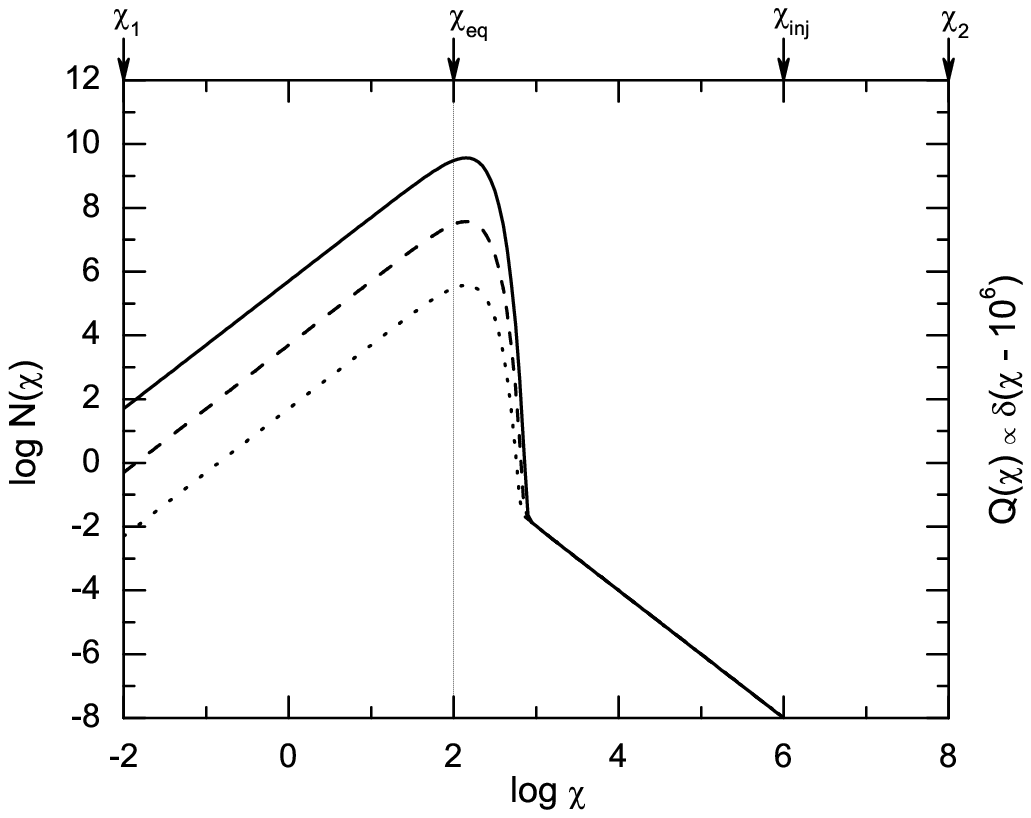}
\caption{The same as Figure (\ref{bohm-syn-1}) except for $\chi_{\rm inj} = 10^6$.}
\label{bohm-syn-2}
\end{figure}

In the case when  $\chi_1 \ll \chi_{\rm eq} \ll \chi_{\rm inj} \ll \chi_2$  the 
asymptotic expansion of the Green's function (\ref{bohm-syn-G}) yields 
\begin{equation}
\left. \mathcal{G}(\chi, \chi_{\rm inj})\right|_{\rm esc, \, \chi_{\rm inj}>}^{q=1} \sim 
\left\{ \begin{array}{ccc}
2^{\eta - 2} \, \Gamma(\eta) \, \chi_{\rm inj}^{-2 \eta} \, \chi_{\rm eq}^{2 \eta -2} \,
\chi^2 \, e^{-{1 \over 2} \, (\chi / \chi_{\rm eq})^2} & {\rm for} & \chi_1 <
\chi \lesssim \chi_{\rm eq} \\
\chi_{\rm inj}^{-2 \eta} \, \chi_{\rm eq}^{2} \, \chi^{2 \eta -2} & {\rm for} & \chi_{\rm
eq} \ll \chi < \chi_{\rm inj} \\
\chi_{\rm inj}^{2 \eta - 4} \, \chi_{\rm eq}^{2} \, \chi^{2\eta -2} \, e^{{1 \over 2} \,
(\chi_{\rm inj} / \chi_{\rm eq})^2} \, e^{-{1 \over 2} \, (\chi / \chi_{\rm eq})^2} & {\rm
for} & \chi_{\rm inj} \lesssim \chi < \chi_2
\end{array} \right. \, .
\label{bohm-G-approx2}
\end{equation}
\noindent
Again as above, the spectrum is different in the case of high energy injection. 
For example, as shown in Figure (\ref{bohm-syn-2}), for the monoenergetic 
injection $Q(\chi) \propto \delta(\chi-\chi_{\rm inj})$ with $\chi_{\rm inj} > \chi_{\rm eq}$ the resulting
electron energy distribution is of the form $N(\chi) \propto \chi^{2} \,
\exp\left[-{1 \over 2} (\chi/\chi_{\rm eq})^2\right]$ for $\chi \lesssim
\chi_{\rm eq}$, while $N(\chi) \propto \chi^{2 \eta - 2}$ for $\chi_{\rm eq} \ll
\chi < \chi_{\rm inj}$. It is interesting to note that the Bohm limit case behaves 
differently from the $q=2$ case and analogous injection condition. The escape timescale
affecs now (via the parameter $\chi_{\rm esc}$, or $\eta$) the
normalization of the low-energy ($\chi <
\chi_{\rm eq}$) segment of the particle spectrum but not its power-law slope. 
It determines, on the other hand,
the `radiatively-cooled' part of the particle distribution in the range
$\chi_{\rm eq} < \chi < \chi_{\rm inj}$, which is, however, very close to the
standard $\propto \chi^{-2}$ for any $\chi_{\rm esc} \gg \chi_{\rm eq}$ (or
$\eta \ll 1$). Here, as before, we
fixed normalization of the monoenergetic injection $\int dp \,
\widetilde{Q}(p)$, and the acceleration and loss timescales, but varied the escape
timescale such that $\chi_{\rm esc} = 10^{5}$, $10^6$, and $10^7$ (dotted,
dashed, and solid lines, respectively). Also we set $\chi_1 = 10^{-2}$,
$\chi_{\rm inj} = 1$, $\chi_{\rm eq} = 10^6$, and $\chi_2 = 10^{8}$.

\section{Emission Spectra}

In the previous sections \S\,3 and \S\,4, we showed that stochastic interactions
of radiating ultrarelativistic electrons (Lorentz factors $\gamma \equiv p /
m_{\rm e} c \gg 1$) with turbulence characterized by a power-law spectrum
$\mathcal{W}(k) \propto k^{-q}$ result in formation of a `universal' high-energy
electron energy distribution
\begin{equation}
n_{\rm e}(\gamma) = n_0 \, \gamma^2 \, \exp\left[- {1\over a} \, \left({\gamma
\over \gamma_{\rm eq}}\right)^a\right] \, ,
\label{ele}
\end{equation}
\noindent
as long as particle escape from the system is inefficient and the radiative
cooling rate scales with some power of electron energy. Here the equilibrium
energy $\gamma_{\rm eq}$ is defined by the balance between acceleration and the
energy losses timescales, while the parameter $a$ depends on the dominant
radiative cooling process and the turbulence spectrum. In particular, for
either synchrotron or IC/Thomson-regime cooling  one has $a = 3 - q$. In the case of dominant IC/KN-regime energy losses
(with $q < 1.5$) one has instead $a = 1.5 - q$. Below we investigate in more
details emission spectra resulting from such an electron distribution.

\subsection{Synchrotron Emission}

Assuming isotropic distribution of momenta of radiating electrons with energy spectrum
$n_{\rm e}(\gamma)$, the synchrotron emissivity can be found as
\begin{equation}
j_{\rm \nu, \, syn}(\nu) = {\sqrt{3} \, e^3 B \over 4 \pi \, m_{\rm e} c^2} \int
\, d\gamma \, \mathcal{R}\!\left({\nu \over \nu_c \, \gamma^2}\right) \, n_{\rm
e}(\gamma) \,
\label{syn-emissivity}
\end{equation}
\noindent
where $\nu_c = 3 e B / 4 \pi m_{\rm e} c$,
\begin{equation}
\mathcal{R}(x) = {x^2 \over 2} \, K_{4/3}\left({x \over 2}\right) \,
K_{1/3}\left({x \over 2}\right) - 0.3 \, {x^3 \over 2} \, \left[
K_{4/3}^2\left({x \over 2}\right) - K_{1/3}^2\left({x \over 2}\right)\right] \,
,
\label{crusius}
\end{equation}
\noindent
and $K_{\mu}\left(z\right)$ is a modified Bessel function of the second order
\citep{cru86}. Relatively complicated function (\ref{crusius}) can be instead
conveniently approximated by $\mathcal{R}(x) \approx 1.81 \times \left( 1.33 +
x^{-2/3} \right)^{-1/2} e^{-x}$ \citep{zir07}, allowing for some analytical
investigation of the integral (\ref{syn-emissivity}). In particular, one may
find that the synchrotron emissivity in a frequency range $\nu < \nu_{\rm syn}
\equiv \nu_c \, \gamma_{\rm eq}^2$ is of the form $j_{\rm \nu, \, syn}(\nu <
\nu_{\rm syn}) \propto \nu^{1/3}$, as expected in the case of a very hard
(inverted) electron energy distribution at low energies, $n_{\rm e}(\gamma < \gamma_{\rm eq})
\propto \gamma^2$. At higher frequencies, however, the synchrotron spectrum
steepens. In order to evaluate such a high-frequency spectral component, we use
the introduced approximation for $\mathcal{R}(x)$, electron spectrum as given in
(\ref{ele}), and with these we re-write synchrotron emissivity
(\ref{syn-emissivity}) as
\begin{equation}
j_{\rm \nu, \, syn}(\nu) \approx {1.81 \, \sqrt{3} \, e^3 B \, \gamma_{\rm eq}^3
n_0 \over 4 \pi \, m_{\rm e} c^2} \int \, dy \, g(\omega, y) \, \exp\left[-
\omega \, h(\omega, y)\right] \, ,
\label{integral}
\end{equation}
\noindent
where $\omega \equiv \nu / \nu_{\rm syn}$, $y \equiv \gamma / \gamma_{\rm eq}$,
$g(\omega, y) \equiv y^2 \, (1.33 + \omega^{-2/3} y^{4/3})^{-1/2}$, and
$h(\omega, y) \equiv y^{-2} + y^a / (\omega \, a)$. 

\begin{figure}
\centering
\includegraphics[scale=1.6]{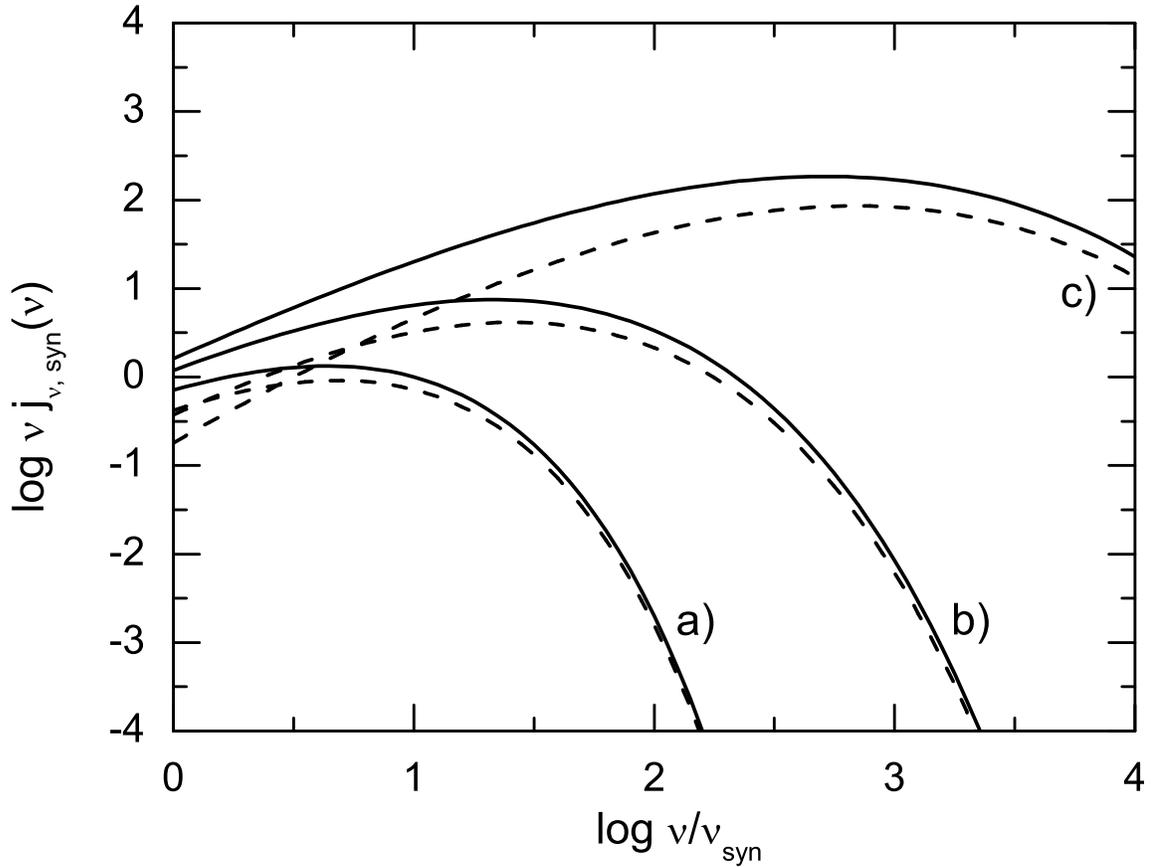}
\caption{Synchrotron spectra resulting from the electron energy distribution
(\ref{ele}) for fixed parameters $B$, $n_0$, and $\gamma_{\rm eq}$. Solid lines
correspond to integration of the exact form of $\mathcal{R}(x)$ as given in
equation (\ref{crusius}), while dashed lines to the approximate formulae
following from (\ref{integral-approx}). Different cases for the parameter $a$
are considered in the plot, namely (a) $a = 3-q$ with $q=1$, (b) $a = 3-q$ with
$q=2$, and (c) $a = 1.5-q$ with $q=1$.}
\label{SYN-SED}
\end{figure}

With such a form it can be
noted that for large $\omega$, i.e. for $\nu > \nu_{\rm syn}$, the integral of
interest can be perform approximately using the steepest descent method \citep[see][]{pet81}. 
This gives
\begin{eqnarray}
& & j_{\rm \nu, \, syn}(\nu>\nu_{\rm syn}) \simeq {1.81 \, \sqrt{3} \, e^3 B \,
\gamma_{\rm eq}^3 n_0 \over 4 \pi \, m_{\rm e} c^2} \, \sqrt{{2 \pi \over \omega
\, h''(\omega, y_{\star})}} \,\, g(\omega, y_{\star}) \, \exp\left[- \omega \,
h(\omega, y_{\star})\right] \simeq
\nonumber \\
& & \simeq {0.54 \, e^3 B \, \gamma_{\rm eq}^3 n_0 \over m_{\rm e} c^2 \, \sqrt{2+a}}
\, \left({2 \nu \over \nu_{\rm syn}}\right)^{{6-a \over 4+2 a}} \, \left[1 +
\left({2 \nu \over \nu_{\rm syn}}\right)^{-{2 a \over 6 + 3 a}}\right]^{-1/2} \,
\exp\left[ - {2 + a \over 2 a} \, \left({2 \nu \over \nu_{\rm syn}}\right)^{{a
\over 2+a}}\right] \, , \label{integral-approx}
\end{eqnarray}
\noindent
where $y_{\star} = (2 \omega)^{1 / (2 +a)}$ is a global maximum of $h(\omega,
y)$, and $h''(\omega, y) = \partial^2 h(\omega, y) / \partial y^2$. Thus, the
high-energy synchrotron component drops much less rapidly than suggested by the
emissivity of a single electron, $\mathcal{R}(x) \propto e^{-x}$. For example,
assuming synchrotron (and/or IC/Thompson-regime) dominance $a = 3-q$, the synchrotron emissivity reads very
roughly as 
\begin{equation}
j_{\rm \nu, \, syn}\left(\nu > \nu_{\rm syn}\right) \propto
\nu^{1/2} \, \exp\left[-1.4 \, \left(\nu / \nu_{\rm syn}\right)^{1/2}\right] \,\,\,\,\,
{\rm for}\,\,\,\,\, q=1,
\end{equation}
or\footnote{As shown by \citet{pet81}, the following spectral form is also true for 
synchrotron emission by semirelativistic electrons.}   
\begin{equation}
j_{\rm \nu, \, syn}\left(\nu > \nu_{\rm syn}\right) \propto
\nu^{5/6} \, \exp\left[-1.9 \, \left(\nu / \nu_{\rm syn}\right)^{1/3}\right]
\,\,\,\,\, {\rm for}\,\,\,\,\,  q=2.
\end{equation} 
In the case of the IC/KN-regime dominance, $a = 1.5 - q$, the
emerging high-energy exponential cut-off in the synchrotron continuum can be
even smoother than this, for example $j_{\rm \nu, \, syn}\left(\nu > \nu_{\rm
syn}\right) \propto \nu^{1.1} \, \exp\left[-2.9 \, \left(\nu / \nu_{\rm
syn}\right)^{0.2}\right]$ for $q=1$. These spectra are shown in Figure
(\ref{SYN-SED}) for fixed parameters $B$, $n_0$, and $\gamma_{\rm eq}$, where
both integration of the exact form of $\mathcal{R}(x)$ as given in equation
(\ref{crusius}) was performed (solid lines), and also approximate formulae
following from (\ref{integral-approx}) were evaluated for comparison (dashed
lines). Different cases for the parameter $a$ are considered in the plot, namely
(a) $a = 3-q$ with $q=1$, (b) $a = 3-q$ with $q=2$, and (c) $a = 1.5-q$ with
$q=1$. As shown, synchrotron spectra are curved and extend far beyond
equilibrium frequency $\nu_{\rm syn}$. In the case of the dominant IC/KN-regime
cooling with $q=1$, the $\nu j_{\nu}(\nu) -
\nu$ synchrotron spectrum peaks  around $\sim 10^3 \nu_{\rm syn}$. We emphasize that the
approximation (\ref{integral-approx}), although obviously not accurate in a
range $\nu \lesssim \nu_{\rm syn}$, works relatively well at higher frequencies,
where the standard $\delta$-approximation for the synchrotron emissivity, $\nu
j_{\rm \nu, \, syn}(\nu) \propto [\gamma^3 \, n_{\rm e}(\gamma)]_{\gamma \propto
\nu^{1/2}}$, fails.

\subsection{Inverse-Compton Emission}

Let us consider inverse-Compton emission of isotropic electrons up-scattering
monoenergetic and isotropic photon field with energy density $u_{\rm ph}$ and
dimensionless photon energy $\epsilon_0 \equiv h\nu_0 / m_{\rm e} c^2$.
The appropriate emissivity can be then found from
\begin{equation}
j_{\rm \nu, \, ic}(\nu) = {3 \, c h \sigma_{\rm T} \over 16 \pi \, m_{\rm e}
c^2} \, u_{\rm ph} \int_{{1 \over 2} \epsilon \, \left(1 + \sqrt{1 + (\epsilon \,
\epsilon_0)^{-1}}\right)} d\gamma \, {\epsilon \over \gamma^2 \epsilon_0^2} \,
\mathcal{J}(\epsilon, \epsilon_0, \gamma) \, n_{\rm e}(\gamma) \, ,
\label{ic-full1}
\end{equation}
\noindent
where $\epsilon \equiv h \nu / m_{\rm e} c^2$, and $\mathcal{J}(\epsilon,
\epsilon_0, \gamma)$ is the IC kernel 
\begin{equation}
\mathcal{J}(\epsilon, \epsilon_0, \gamma) = 2 \, \mathcal{I} \, \ln \mathcal{I}
+ \mathcal{I} + 1 - 2 \, \mathcal{I}^2 + {\mathcal{L}^2 \mathcal{I}^2 \, (1
-\mathcal{I}) \over 2 \, (1 - \mathcal{L} \, \mathcal{I})} \quad {\rm with}
\quad \mathcal{L} \equiv 4 \epsilon_0 \gamma \quad {\rm and} \quad \mathcal{I}
\equiv {\epsilon \over \mathcal{L} \, (\gamma - \epsilon)} 
\label{ic-kernel} 
\end{equation}
\noindent
\citep[e.g.,][]{blu70}. 

\begin{figure}
\centering
\includegraphics[scale=1.6]{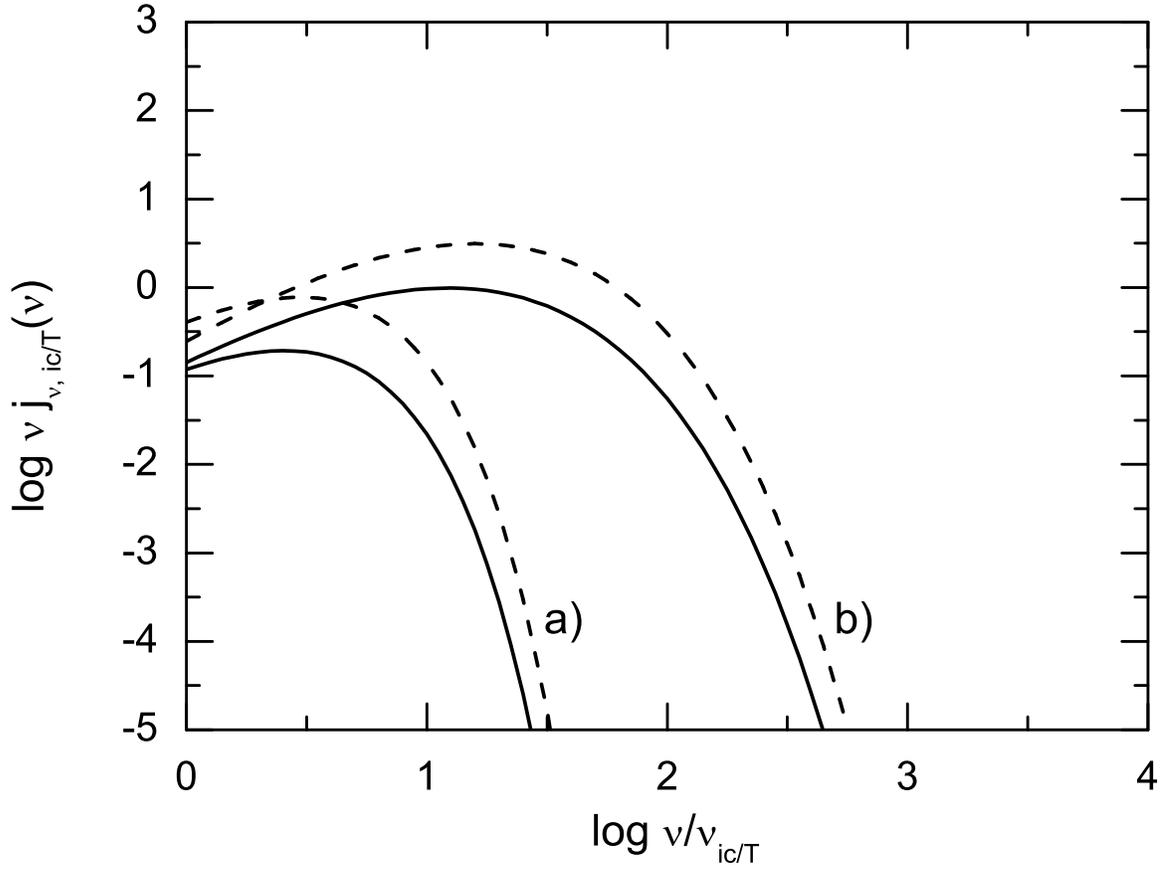}
\caption{Inverse-Compton spectra produced in the Thomson regime, resulting from
the electron energy distribution (\ref{ele}) for fixed parameters $B$, $n_0$,
and $\gamma_{\rm eq}$. Solid lines correspond to the formulae (\ref{ic-full2}),
and dashed lines to the rough approximation (\ref{thom-high}). Two different
parameters $a = 3-q$ are considered in the plot, corresponding to the turbulence
energy index $q=1$ and $q=2$ (cases (a) and (b), respectively).}
\label{IC-T-SED}
\end{figure}

\begin{figure}
\centering
\includegraphics[scale=1.6]{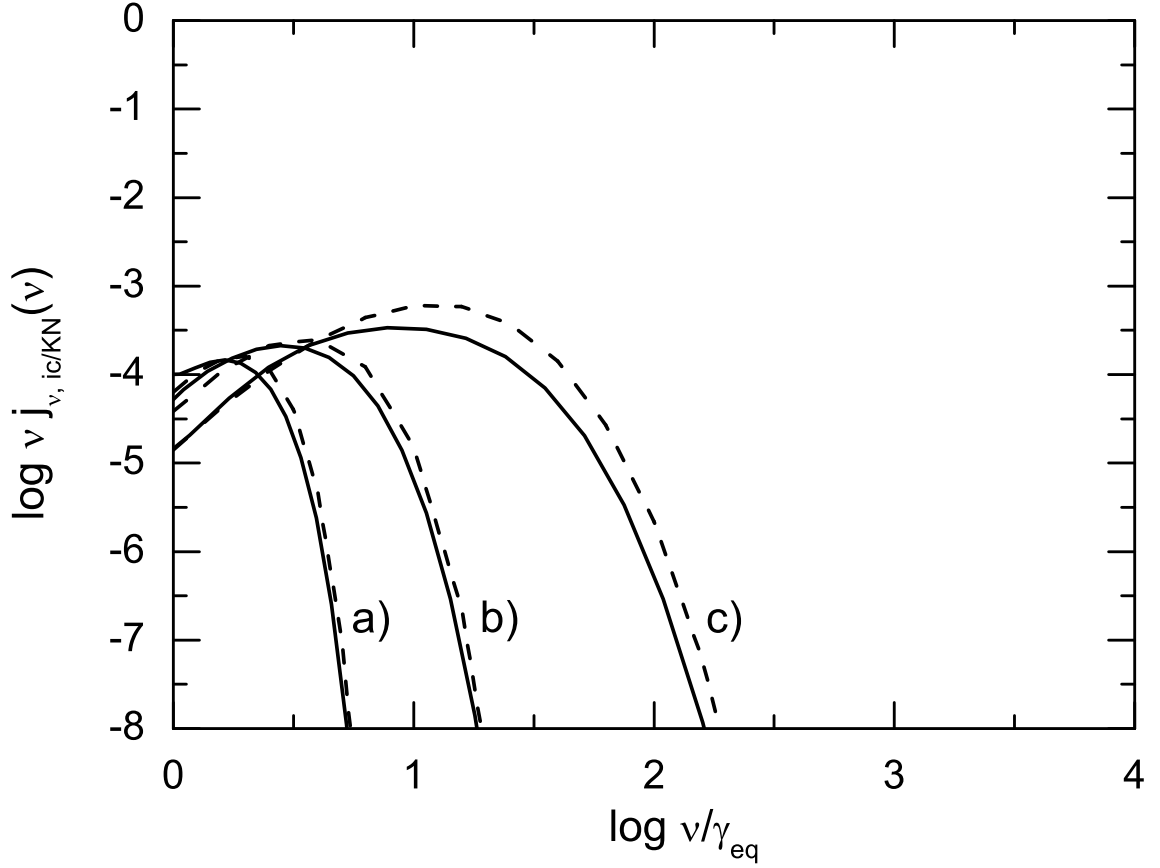}
\caption{Inverse-Compton spectra produced in the KN regime, resulting from the
electron energy distribution (\ref{ele}) for fixed parameters $B$, $n_0$, and
$\gamma_{\rm eq}$. Solid lines correspond to the exact evaluation of the
integral (\ref{ic-full1}), and dashed lines to the rough approximation
(\ref{rough-KN}). Different cases for the parameter $a$ are considered in the
plot, namely (a) $a = 3-q$ with $q=1$, (b) $a = 3-q$ with $q=2$, and (c) $a =
1.5-q$ with $q=1$. For illustration $\gamma_{\rm cr}/\gamma_{\rm eq} = 0.01$ has
been selected.}
\label{IC-KN-SED}
\end{figure}

Let us discuss first the case when the KN effects are negligible. The IC kernel
can then be approximated by $\mathcal{J}(\epsilon, \epsilon_0, \gamma) \approx
{2 \over 3} (1-\omega / y^2)$, with $y \equiv \gamma / \gamma_{\rm eq}$, $\omega
\equiv \epsilon / \epsilon_{\rm ic/Th}$, and $\epsilon_{\rm ic/Th} \equiv 4
\epsilon_0 \, \gamma_{\rm eq}^2$ which is the characteristic energy of soft photon
inverse-Compton up-scattered in a Thomson regime by electrons with Lorentz
factor $\gamma_{\rm eq}$. Hence, with the electron energy distribution of the
form (\ref{ele}), one can find that
\begin{eqnarray}
& & \epsilon j_{\rm \epsilon, \, ic/Th}(\epsilon) = {2 \over \pi} \, c \sigma_{\rm Th}
\, u_{\rm ph} \, n_0 \, \gamma_{\rm eq}^5 \, \int_{\sqrt{x}} \, dy \, \omega^2 \,
\left( 1 - {\omega \over y^2}\right) \, \exp\!\left[-{1 \over a} \, y^a\right] \approx
\nonumber \\
& & \approx {2 \over \pi} \, c \sigma_{\rm T} \, u_{\rm ph} \, n_0 \, \gamma_{\rm eq}^5 \,
a^{-1} \omega^2 \, \left\{ a^{1/a} \, \Gamma\!\left[a^{-1}, a^{-1} \,
\omega^{a/2}\right] - a^{-1/a} \, \omega \, \Gamma\!\left[- a^{-1}, a^{-1} \,
\omega^{a/2}\right]\right\} \, , \label{ic-full2}
\end{eqnarray}
\noindent
where $\Gamma[a,z]$ is incomplete Gamma function. With the expansion
$\Gamma[a,z] \sim \Gamma[a]$ for $z \rightarrow 0$ \citep{abr64}, one can
approximate further
\begin{equation}
\epsilon j_{\rm \epsilon, \, ic/Th}(\epsilon < \epsilon_{\rm ic/Th}) \sim {2 \over
\pi} \, c \sigma_{\rm Th} \, u_{\rm ph} \, n_0 \, \gamma_{\rm eq}^5 \, a^{{1-a \over a}}
\, \Gamma\!\left(a^{-1}\right) \, \left({\epsilon \over \epsilon_{\rm
ic/Th}}\right)^2 \, .
\label{thom-low}
\end{equation}
\noindent
In other words, the IC emissivity at low photon energies is of the form $j_{\rm
\epsilon, \, ic/Th}(\epsilon < \epsilon_{\rm ic/Th}) \propto \epsilon$. This is
the flattest IC/Thomson-regime spectrum, being analogous to the flattest
synchrotron one $j_{\rm \nu, \, syn}(\nu < \nu_{\rm syn}) \propto \nu^{1/3}$. At
higher photon energies, noting that $\Gamma[a,z] \sim z^{a-1} \, e^{-z}$ for $z
\rightarrow \infty$, one may find instead
\begin{equation}
\epsilon j_{\rm \epsilon, \, ic/Th}(\epsilon > \epsilon_{\rm ic/Th}) \sim {2 \over
\pi} \, c \sigma_{\rm Th} \, u_{\rm ph} \, n_0 \, \gamma_{\rm eq}^5 \, \left({\epsilon
\over \epsilon_{\rm ic/Th}}\right)^{{5-a \over 2}} \, \exp\left[- {1 \over a} \,
\left({\epsilon \over \epsilon_{\rm ic/Th}}\right)^{{a \over 2}} \right] \, .
\label{thom-high}
\end{equation}
\noindent
Therefore, the exponential cut-off of the IC/Thomson-regime component is now
steeper than the exponential cut-off of the synchrotron component originating
from the same particle distribution. In particular, with $a = 3-q$ one gets
$j_{\rm \epsilon, \, ic/Th}(\epsilon > \epsilon_{\rm ic/Th}) \propto
\epsilon^{1/2} \, \exp[ - {1 \over 2} \, (\epsilon / \epsilon_{\rm ic/Th})]$ for
$q=1$, while $j_{\rm \epsilon, \, ic/Th}(\epsilon > \epsilon_{\rm ic/Th}) \propto
\epsilon \, \exp[ - (\epsilon / \epsilon_{\rm ic/Th})^{1/2}]$ for $q=2$ (that can
be compared with the corresponding synchrotron emissivities provided above).
These spectra are shown in Figure (\ref{IC-T-SED}) for fixed parameters $B$,
$n_0$, and $\gamma_{\rm eq}$. Here the solid lines correspond to the formulae
(\ref{ic-full2}), and dashed lines to the rough approximation (\ref{thom-high}).
Two different parameters $a = 3-q$ are considered in the plot, corresponding to
the turbulence energy index $q=1$ and $q=2$ (cases (a) and (b), respectively).

Finally, we comment on the emission spectra produced in a deep KN regime of the
IC scattering, i.e. when $\gamma > \gamma_{\rm cr} \equiv 1/ 4 \epsilon_0$, by
the highest-energy electrons $\gamma \gtrsim \gamma_{\rm eq}$. In such a case,
the emissivity has to be evaluated by performing the integral (\ref{ic-full1})
with the exact IC kernel as given in equation (\ref{ic-kernel}). A rather crude
approximation for such can be obtained by utilizing the $\delta$-approximation
for the resulting IC/KN-regime photon energy, namely $\epsilon = \gamma$. In
particular, with the electron energy distribution as given in (\ref{ele}), and
with all the previous assumptions regarding monoenergetic and isotropic soft
photon field, one finds
\begin{eqnarray}
& & \epsilon j_{\rm \epsilon, \, ic/KN}(\epsilon \gtrsim \gamma_{\rm eq}) \simeq
{m_{\rm e} c^2 \over 4 \pi}\left. {\gamma^2 \, n_{\rm e}(\gamma) \over t_{\rm
IC}(\gamma)}\right|_{\gamma = \epsilon} \simeq \nonumber \\
& & \simeq {1 \over 3 \pi} \, c \sigma_{\rm T} \, u_{\rm ph} \, n_0 \, \gamma_{\rm eq}^5 \,
\left({\epsilon \over \gamma_{\rm eq}}\right)^5 \, \left(1 + {\epsilon \over
\gamma_{\rm eq}} \, {\gamma_{\rm eq} \over \gamma_{\rm cr}}\right)^{-1.5} \,
\exp\left[- { 1 \over a} \, \left({\epsilon \over \gamma_{\rm
eq}}\right)^a\right] \, ,
\label{rough-KN}
\end{eqnarray}
\noindent
where $t_{\rm IC}(\gamma)$ is the inverse-Compton cooling timescale as
introduced previously in equation (\ref{ic}). As shown in Figure \ref{IC-KN-SED}, as a result the IC/KN-regime spectra
cut-off sharply above $\epsilon = \gamma_{\rm eq}$ photon energies, imitating
exponential cut-off in the energy distribution of radiating particles. Here the exact calculations are plotted as solid lines, and
rough approximation (\ref{rough-KN}) as dashed ones. We fix parameters $B$, $n_0$, 
$\gamma_{\rm eq}$, $\gamma_{\rm cr}/\gamma_{\rm eq} = 0.01$, and  again, different cases for
the parameter $a$ are considered; (a) $a = 3-q$ with $q=1$, (b) $a = 3-q$
with $q=2$, and (c) $a = 1.5-q$ with $q=1$. We also choose for illustration
$\gamma_{\rm cr}/\gamma_{\rm eq} = 0.01$.

\section{Discussion and Conclusions}

In this paper we study steady-state spectra of ultrarelativistic electrons
undergoing momentum diffusion due to resonant interactions with turbulent MHD
waves. We assume a given power spectrum $\mathcal{W}(k) \propto k^{-q}$ for
magnetic turbulence within some finite range of turbulent wavevectors $k$, and
consider variety of turbulence spectral indices $1 \leq q \leq 2$. For example,
$q=1$ corresponds to the `Bohm limit' of the stochastic acceleration processes,
$q = 2$ represents the `hard-sphere approximation', while $q = 5/3$ and $q=3/2$
to the Kolmogorov or Kreichnan turbulence, respectively. Within the anticipated
quasilinear approximation for particle-wave interactions, such a turbulent
spectrum gives the momentum and pitch angle diffusion rates $\propto p^{q-2}$, or the
acceleration and escape timescales $t_{\rm acc} \propto p^{2 - q}$ and $t_{\rm
esc} \propto p^{q-2}$. In the analysis, we also include radiative energy losses,
being an arbitrary function of the electrons' energy. In most of the cases,
however, or at least in some particular energy ranges, the appropriate timescale
for the radiative cooling scales simply with some power of the particle
momentum, $t_{\rm loss} \propto p^r$. For example, $r = -1$ corresponds to
synchrotron or inverse-Compton/Thomson-regime energy losses, $r=0$ (roughly) to
the bremsstrahlung emission, $r = +1$ (roughly) to the Coulomb interactions of
ultrarelativistic electrons, while $r = 1/2$ may conveniently approximate
inverse-Compton cooling in the Klein-Nishina regime on monoenergetic 
background soft photon
field. 

We find that when the particles are confined to the turbulent acceleration region 
($t_{\rm esc}\rightarrow\infty$), the resulting steady-state particle spectra 
(for a finite momentum range of interacting electrons) are in general of the
modified ultrarelativistic Maxwellian type, $n_{\rm e}(p) \propto p^2 \,
\exp\left[ - {1 \over a} \, \left(p / p_{\rm eq}\right)^a\right]$, where $a =
2-q-r \neq 0 $. Here $p_{\rm eq}$ is the momentum at which the 
acceleration and radiative loss timescales are equal, $t_{\rm
acc}(p_{\rm eq}) = t_{\rm loss}(p_{\rm eq})$. This form is independent of 
the initial energy distribution of the electrons as long as this distribution is 
not very broad and the bulk of initial particles have $p<p_{\rm eq}$. However, 
if high energy particles with $p>p_{\rm eq}$ are injected to the system, there 
will be significant deviations from this simple form. For example, for a $\delta$-function 
initial distribution the spectrum will have a power-law tail $\propto
p^{r-1}$ in addition to the modified Maxwellian bump. Also, if
the ratio of acceleration and energy losses timescales is independent of the
electron energy, in other words, if $2-q = r$, then the resulting particle spectra
are of the form $n_{\rm e}(p) \propto p^{-\sigma'}$, where $\sigma' \equiv
(t_{\rm acc}/t_{\rm loss}) - 2$. Finally, if the particle escape from the
acceleration site is finite but still inefficient, a power-law tail $\propto
p^{1-q}$ may be present in the momentum range $p_{\rm inj} < p \ll p_{\rm eq}$,
again in addition to the modified Maxwellian component. When the
radiative losses timescale is not a simple power-law function of the electron
energy, the emerging spectra may be of a more complex (e.g., concave) form. 

We also analyze in more details synchrotron and inverse-Compton emission spectra
of the electrons characterized by the modified ultrarelativistic Maxwellian
energy distribution. In order to summarize briefly our findings, let us define
the critical synchrotron frequency of the electrons with the equilibrium Lorentz
factor $\gamma_{\rm eq} \equiv p_{\rm eq}/m_{\rm e} c$, namely $\nu_{\rm syn}
\equiv (3 e B / 4 \pi m_{\rm e} c) \, \gamma_{\rm eq}^2$, and the critical
dimensionless energy of the monochromatic ($h\nu_0 \equiv \epsilon_0 \, m_{\rm
e}c^2$) soft photon field inverse-Compton up-scattered (in the Thomson
regime) by the $\gamma_{\rm eq}$ electrons, $\epsilon_{\rm ic/Th} = 4 \,
\epsilon_0 \, \gamma_{\rm eq}^2$. With these, one can note that the
low-frequency synchrotron emissivity is of the form $j_{\rm \nu, \, syn}(\nu <
\nu_{\rm syn}) \propto \nu^{1/3}$, as expected in the case of a very flat
(or inverted) electron energy distribution $n_{\rm e}(\gamma < \gamma_{\rm eq})
\propto \gamma^2$. Such flat electron spectra seem to be required to explain
several emission properties of relativistic jets in active galactic nuclei
\citep{tsa07a,tsa07b}. At higher frequencies, we find a rough approximation
$j_{\rm \nu, \, syn}(\nu>\nu_{\rm syn}) \propto \nu^{(6-a) / (4+2 a)} \,
\exp\left[ - {2 + a \over 2 a} \, \left(2 \nu / \nu_{\rm syn}\right)^{a /
(2+a)}\right]$. Thus, the high-energy synchrotron component drops much less
rapidly than suggested by the emissivity of a single electron, and the emerging
high-frequency tail of the synchrotron spectrum is of a smoothly curved shape.
It is therefore very interesting to note that almost exactly this kind of
curvature is observed at synchrotron X-ray frequencies in several BL Lac objects
\citep{mas04,mas06,per05,tra07a,tra07b,gie07}, in particular those detected also
at TeV photon energies. 

As for the inverse-Compton emission of ultrarelativistic electrons characterized
by the modified Maxwellian energy distribution, we find that in the Thomson
regime it is of the form $j_{\rm \epsilon, \, ic/Th}(\epsilon < \epsilon_{\rm
ic/Th}) \propto \epsilon$, and $j_{\rm \epsilon, \, ic/Th}(\epsilon >
\epsilon_{\rm ic/Th}) \propto \epsilon^{(3-a) / 2} \, \exp\left[ - {1 \over a} \,
\left(\epsilon / \epsilon_{\rm ic/Th}\right)^{a / 2}\right]$. Both very flat
low-energy part of this component and also its curved high-energy segment may
contribute to the observed $\gamma$-ray emission of some TeV blazars
\citep{kat06a,gie07}\footnote{The caution here is that the computed in this paper 
high-energy spectra correspond to the situation of inverse-Comptonization of the
monoenergetic seed photon field, which, in addition, is isotropically distributed
in the emitting region rest frame. In the case of relativistic blazar jets, the 
external radiation (due to accretion disk, as well as circumnuclear gas and dust)
is distributed anisotropically in the jet rest frame, while the isotropic 
synchrotron emission produced by the jet electrons is not strictly monochromatic
\citep[see, e.g.,][and references therein]{der97}. 
One the other hand, synchrotron radiation of ultrarelativistic electrons 
characterized by the Maxwellian-type energy distribution, as analyzed here, is
not that far from the monoenergetic approximation, and the relativistic corrections
regarding the anisotropic distribution of the soft photons in the emitting region 
rest frame are not supposed to influence substantially the spectral shape of the 
inverse-Compton emission. For these reasons, we believe that the main spectral 
features of the high-energy emission components computed in this paper are 
representative for the $\gamma$-ray emission of, e.g., TeV blazars.}. 
We also note, that the curvature of the high frequency
segments of the synchrotron and inverse-Compton spectra, event though being
produced by the same energy electrons and in the Thomson regime, are different.
Such a difference is even more pronounce when the Klein-Nishina effects play a
role, since in such a case an exponential decrease of the high-energy photon
spectra is the strongest, $j_{\rm \epsilon, \, ic/KN}(\epsilon > \gamma_{\rm
eq}) \propto \epsilon^{7/2} \, \exp\left[ - {1 \over a} \, \left(\epsilon /
\gamma_{\rm eq}\right)^{a}\right]$, imitating exponential cut-off in the energy
distribution of radiating particles.

\acknowledgments

\L .S. was supported by MEiN through the research project 1-P03D-003-29 in years
2005-2008. \L .S. acknowledges M. Ostrowski, R. Schlickeiser, and S. Fuerst for
helpful comments and discussion.

\end{document}